\documentclass[aps,prc,twocolumn,superscriptaddress,nofootinbib,longbibliography,floatfix,10pt]{revtex4-2}

\usepackage{graphicx}
\usepackage{dcolumn}
\usepackage{bm}
\usepackage{morefloats}
\usepackage{multirow}
\usepackage{amssymb}
\usepackage{amsmath}
\usepackage{xcolor}
\usepackage{longtable}
\usepackage{fix-cm}
\usepackage{verbatim}
\usepackage{mathptmx} 
\usepackage[T1]{fontenc}
\usepackage[colorlinks,allcolors=blue]{hyperref}

\makeatletter
\def\NAT@def@citea{\def\@citea{\NAT@separator}}
\makeatother

\begin{document}

\title{Neutron-rich isotope production for  $Z\geq 98$ in ${}^{238} \mathrm{U}+{ }^{248} \mathrm{Cm}$ reaction}

\author{S.E. Ocal}
\affiliation{Physics Department, Middle East Technical University, 06800 Ankara, Turkey}
\author{O. Yilmaz}
\affiliation{Physics Department, Middle East Technical University, 06800 Ankara, Turkey}
\author{S. Ayik}\email{ayik@tntech.edu}
\affiliation{Physics Department, Tennessee Technological University, Cookeville, TN 38505, USA}
\author{A. S. Umar}
\affiliation{Department of Physics and Astronomy, Vanderbilt University, Nashville, TN 37235, USA}

\date{\today}

\begin{abstract}
\edef\oldrightskip{\the\rightskip}
\begin{description}
\rightskip\oldrightskip\relax
\setlength{\parskip}{0pt} 
\item[Background] Multi-nucleon transfer (MNT) reactions in actinide systems are
a promising method to synthesize transuranium neutron-rich elements. Appropriate
theoretical approaches are needed to understand the mechanism behind MNT.
\item[Purpose] This work aims to produce neutron-rich isotopes in the super-heavy region
through the ${}^{238} \mathrm{U}+{ }^{248} \mathrm{Cm}$  system. We employ a microscopic approach
to elucidate reaction mechanisms, and predict new isotope production that expands the known nuclear chart. 
\item[Methods] The stochastic mean-field (SMF) approach, including fluctuations and correlations, is used to
explain the primary cross-sections in MNT reactions based on the quasi-fission and inverse quasi-fission
processes, and a statistical de-excitation model with GEMINI++ code to calculate
the secondary fragment cross-sections.
\item[Results] The calculated cross-sections using SMF and GEMINI++ explain available
experimental results for the ${}^{238} \mathrm{U}+{ }^{248} \mathrm{Cm}$  system at $E_\mathrm{c.m.}=898.7$~MeV energy. This
shows the effectiveness and applicability of the quantal diffusion approach based on
the SMF theory in heavy-ion collisions.
\item[Conclusions]
Production of transuranium neutron-rich elements with a proton
number up to $Z=$101 are obtained with sizable cross-sections. Theoretical results calculated for the
Z=102-105 region, for which there are no experimental data, show that the cross-section
values would be lower than the microbarn level. SMF theory does not contain any
adjustable parameters other than the standard parameters of the energy density
functional used in the TDHF theory and is an important approach for the microscopic
understanding of reaction mechanisms.
\end{description}
\end{abstract}

\maketitle

\section{Introduction}\label{sec_1}

Extensive theoretical and experimental research is underway to produce heavy
elements in the region between the presently known isotopes and the
driplines, particularly in the upper half of the nuclear chart on the neutron-rich side. 
Production of the unknown isotopes closed to the neutron and proton drip
lines will give new input for the models, particularly new spherical
and deformed shell effects in the region of super-heavy nuclei.
Fusion is presently the only method to synthesize transuranium and superheavy
nuclei, but the measured production cross-sections are expected to be very small,
below picobarns, requiring an increase in beam intensities and improvement in
experimental techniques for isotope separation and detection~\cite{adamian2020,munzenberg2023}.
Multi-nucleon transfer (MNT) reactions that occur in deep-inelastic binary collisions near the Coulomb barrier
energies are an alternate method for producing neutron-rich heavy isotopes using actinide
targets, and is being experimentally studied in laboratories around the world~\cite{niwase2023}.

Theoretically, several phenomenological techniques, such as the multidimensional Langevin model
\cite{zagrebaev2008,zagrebaev2008c,zagrebaev2011,karpov2017,saiko2019,saiko2022}, the di-nuclear system model~\cite{feng2009a,feng2017}, and the quantum molecular dynamics
model~\cite{zhao2009,zhao2016,wang2016} have been used to study the multinucleon transfer mechanism. The time-dependent Hartree-Fock (TDHF) theory provides a microscopic description of reaction
dynamics and has been widely used to analyze MNT reactions~\cite{simenel2012,simenel2018,nakatsukasa2016,oberacker2014,umar2015a,umar2015c,umar2017,simenel2010,sekizawa2016,godbey2019,simenel2020}.
Beyond the mean-field approximation of TDHF, the time-dependent random phase approximation (TDRPA), developed by Balian and V\'en\'eroni,
is used for symmetric systems~\cite{balian1985,balian1992,simenel2011,williams2018,godbey2020,godbey2020b}. The
stochastic mean-field (SMF) approach, which includes mean-field fluctuations and
correlations between proton and neutron transfers, provides an additional improvement
to the TDHF theory, where quantal effects, memory effects, and the full collision geometry are
included~\cite{ayik2008,lacroix2014}. In the SMF framework, the production cross sections of neutron-rich isotopes based on quasi-fission reactions (QF) and inverse quasi-fission
reactions (IQF) in MNT reactions can be calculated~\cite{ayik2018,ayik2019,sekizawa2020,yilmaz2020,ayik2020,ayik2021,ayik2023,ayik2023b,kayaalp2024}. This work presents our theoretical results for MNT reactions of the  ${}^{238} \mathrm{U}+{ }^{248} \mathrm{Cm}$ system
at $E_\mathrm{c.m.}=898.7$~MeV energy by employing the SMF approach to synthesize neutron-rich nuclei above $Z\geq 98$. Here, ${ }^{238} \mathrm{U}+{ }^{248} \mathrm{Cm}$ is preferred as one of the heaviest and most
neutron-rich available target materials. In addition, the excited primary products are
cooled down using the statistical de-excitation GEMINI++ code and the
production cross-sections for the secondary fragments are calculated. The ${}^{238} \mathrm{U}+{ }^{248} \mathrm{Cm}$ system at
$E_\mathrm{c.m.}=898.7$~MeV energy has been studied both experimentally~\cite{kratz2013,schaedel1982,kratz2015} and
theoretically~\cite{peng2022,zhao2023,zagrebaev2008}. In this work, the production cross sections of the primary and
secondary products in the transuranium region (such as Cf, Es, Fm, Md) are
calculated by choosing four different initial orientations of projectile and target nuclei
and compared with experimental data. We also present the calculation for $Z=$102-105
(No, Lr, Rf, and Db) for which there are no experimental data. We briefly discuss the
theoretical framework of the cross-section calculations for reaction products in Sec. II.
In Sec. III, we present our calculations for the ${}^{238} \mathrm{U}+{ }^{248} \mathrm{Cm}$ 
system at $E_\mathrm{c.m.}=898.7$~MeV energy using the quantal transport description based on the SMF approach.
Conclusions are presented in Sec. IV.

\section{Theoretical Framework}\label{sec_2}
\subsection{Primary cross-sections of reaction products}\label{sec_2.1}
In quasifission reactions, the colliding ions stick and move together for an extended time compared to typical reaction times,
during this time they exchange nucleons without fusing, they then
separate into the primary di-nuclear system with different numbers of neutrons and
protons than the initial ones. We calculate the cross-sections for the production of
primary isotopes using the standard expression,

\begin{align}\label{eq20}
\sigma ^{pri}(N,Z)=\frac{\pi \hbar ^2}{2{\mu}E_\mathrm{c.m.}}\sum \limits_{\ell_{min}}^{\ell_{max}} (2\ell +1 )\frac{1}{2}[P^{pro}_{\ell}(N,Z)+P^{tar}_{\ell}(N,Z)]
\end{align}
In this expression,
$P^{pro}_{\ell}(N,Z)$
and
$P^{tar}_{\ell}(N,Z)$
denote the normalized probability
distribution of producing projectile-like and target-like fragments. The factor of 1/2 is
introduced to normalize the total distribution of the primary fragment to unity. A
Fokker-Planck type equation~\cite{risken1996} describes the evolution of the probability distribution
function of macroscopic variables, which is a correlated Gaussian function of the mean
values of neutron, proton, and mixed dispersions 
\begin{equation}
P_{\ell}(N, Z,t)=\frac{1}{2 \pi \sigma_{N N}(\ell) \sigma_{Z Z}(\ell) \sqrt{1-\rho_{\ell}^{2}}} \exp \left(-C_{\ell}\right)\;. \label{Eq:Jprob}
\end{equation}
Here, the exponent $C_l$ for each initial angular momentum is given by 
\begin{equation}
C_{\ell} = \frac{1}{2\left(1-\rho_{\ell}^{2}\right)} \begin{aligned}[t]
& \left[\left(\frac{Z-Z_{\ell}}{\sigma_{Z Z}(\ell)}\right)^{2} 
- 2\rho_{\ell} \left(\frac{Z-Z_{\ell}}{\sigma_{Z Z}(\ell)}\right)\left(\frac{N-N_{\ell}}{\sigma_{N N}(\ell)}\right) \right. \\
& \left. + \left(\frac{N-N_{\ell}}{\sigma_{N N}(\ell)}\right)^{2}\right]\;,
\end{aligned}\;,
\end{equation}
with the correlation coefficient $\rho_\ell=\sigma_{NZ}^2(\ell)/\sigma_{ZZ}(\ell)\sigma_{NN}(\ell)$.  Quantities $N_\ell$ and
$Z_\ell$ denote the mean neutron and proton numbers of the target-like and projectile-like
fragments that are determined by the TDHF calculations. Dispersions satisfy a set of coupled
differential equations~\cite{schroder1981,merchant1981}

\begin{equation}
\frac{\partial}{\partial t } {\sigma}^2_{NN} = 2 \frac{\partial
\nu_{n}}{\partial N_1} \sigma^2_{NN} + 2 \frac{\partial \nu_{n}}{\partial Z_1}\sigma^2_{NZ} + 2 D_{NN}\;, \label{Eq:Coupled1}
\end{equation}
\begin{equation}
\frac{\partial}{\partial t } {\sigma}^2_{ZZ} = 2 \frac{\partial
\nu_{p}}{\partial Z_1} \sigma^2_{ZZ} + 2 \frac{\partial \nu_{p}}{\partial
N_1}\sigma^2_{NZ} + 2 D_{ZZ}\;,
\label{Eq:Coupled2}
\end{equation}
\begin{equation}
\frac{\partial}{\partial t } {\sigma}^2_{NZ} =  \frac{\partial \nu_{p}}{\partial
N_1} \sigma^2_{NN} +  \frac{\partial \nu_{n}}{\partial Z_1}\sigma^2_{ZZ} +
\sigma^2_{NZ}\left(\frac{\partial \nu_p}{\partial Z_1} +\frac{\partial
\nu_n}{\partial N_1}\right)\; \label{Eq:Coupled3}
\end{equation}
To determine neutron, proton, and mixed dispersions we need to calculate the
diffusion coefficients ($D_{NN}$ and $D_{ZZ}$) of neutrons and protons, and the derivatives of the drift
coefficients ($v_p(t)$
and $v_n(t)$).

\subsection{Quantal diffusion coefficients}\label{sec_2.2}
In the nucleon diffusion mechanism, the neutron number and proton number of the
projectile-like or target-like fragments are the relevant macroscopic variables. For each
stochastically generated event $\lambda$ in an ensemble, the neutron and proton numbers, $N_1^{\lambda}(t)$ and $Z_1^{\lambda}(t)$, are
determined by integrating the nucleon density over the projectile side of the window
between the colliding nuclei. Based on SMF theory, the rate of changes of neutron and proton numbers $dN_1(t)/dt$ and $dZ_1(t)/dt$ satisfy a coupled set of Langevin equations.  We consider the small amplitude
fluctuations and linearize the Langevin equation around the mean values of the
macroscopic variables $N_1^{\lambda}(t)=N_1(t)-\delta N_1^{\lambda}(t)$ and $Z_1^{\lambda}(t)=Z_1(t)-\delta Z_1^{\lambda}(t)$. The mean
values $N_1(t)$ and $Z_1(t)$ are determined by the mean-field description of TDHF theory
for small amplitude fluctuations. The fluctuations of the neutron and proton numbers $\delta N_1^{\lambda}$, $\delta Z_1^{\lambda}$
evolve according to coupled linear quantal Langevin equations~\cite{risken1996,schroder1981,merchant1981,gardiner1991,weiss1999,norenberg1981}, 
\begin{equation}
\begin{split}
\frac{d}{d t}\left(\begin{array}{l}
\delta \mathrm{Z}_{1}^{\lambda} \\
\delta N_{1}^{\lambda}
\end{array}\right)=& \left(\begin{array}{l}
\frac{\partial v_{p}}{\partial Z_{1}}\left(Z_{1}^{\lambda}-\overline{Z_{1}}\right) +\frac{\partial v_{p}}{\partial N_{1}}\left(N_{1}^{\lambda}-\overline{N_{1}}\right) \\
\frac{\partial v_{n}}{\partial Z_{1}}\left(Z_{1}^{\lambda}-\overline{Z_{1}}\right) +\frac{\partial v_{n}}{\partial N_{1}}\left(N_{1}^{\lambda}-\overline{N_{1}}\right)
\end{array}\right) \\
&\quad \quad \quad \quad+\left(\begin{array}{l}
\delta v_{p}^{\lambda}(t) \\
\delta v_{p}^{\lambda}(t)
\end{array}\right)\;.
\end{split}
\label{Eq:CoupLang}
\end{equation}
Langevin equation is equivalent to the Fokker-Plank equation for the distribution
function of the macroscopic variables. The stochastic part of the drift coefficients $\delta v^\lambda_{n,p}(t)$, which provides the source for generating fluctuations in mass and charge
asymmetry degrees of freedom, are determined by uncorrelated Gaussian
distributions with zero mean values $\overline{\delta} v^\lambda_{n,p}(t)=0$. The integration of these autocorrelation functions over history defines the diffusion coefficients $D_{NN}(t)$
for proton
and neutron transfers~\cite{gardiner1991,weiss1999}
\begin{align}
\int_{0}^{t}dt'\overline{\delta v_{\alpha }^{\lambda } (t)\delta v_{\alpha }^{\lambda}(t')} =D_{\alpha \alpha } (t)\;.
\end{align}
Diffusion coefficients generally involve a complete set of particle-hole states~\cite{ayik2018}. It
is possible to eliminate the entire set of particle states by employing closure relations
in a diabatic limit~\cite{norenberg1981}. This results in an important simplification, and as a result,
diffusion coefficients are determined entirely in terms of the occupied single-particle
wave functions of TDHF evolution. Explicit expressions of diffusion coefficients are
provided in previous publications~\cite{ayik2018,ayik2019,sekizawa2020}. The fact that diffusion coefficients are
determined by the mean-field properties is consistent with the fluctuation-dissipation
theorem of non-equilibrium statistical mechanics and it greatly simplifies calculations
of quantal diffusion coefficients. Diffusion coefficients include the quantal effects due
to shell structure, Pauli blocking, and the full effect of the collision’s geometry without
any adjustable parameters. The direct part is given as the sum of the nucleon currents
across the window from the target-like fragment to the projectile-like fragment and from the projectile-like fragment to the target-like fragment, which is integrated over the
memory. This is analogous to the random walk problem, in which the diffusion
coefficient is given by the sum of the rate for forward and backward steps~\cite{weiss1999,norenberg1981,randrup1979}. The second part in the quantal diffusion expression stands for the Pauli blocking
effects in the nucleon transfer mechanism, which does not have a classical
counterpart~\cite{ayik2018},

\begin{align}
\label{eq.1}
D_{\alpha \alpha}(t) &= \int_0^t d\tau \int d^3 r \, g(x') \left[ G_T(\tau) J_{\perp \alpha}^T\left(\vec{r}, t - \frac{\tau}{2}\right) \right. \nonumber \\
&\quad + \left. G_P(\tau) J_{\perp \alpha}^P\left(\vec{r}, t - \frac{\tau}{2}\right) \right] \nonumber \\
&\quad - \int_0^t d\tau \, \text{Re} \Bigg[ \sum_{h' \in P, h \in T} A_{h' h}^\alpha(t) A_{h' h}^{\alpha *}(t - \tau) \nonumber \\
&\quad + \sum_{h' \in T, h \in P} A_{h' h}^\alpha(t) A_{h' h}^{\alpha *}(t - \tau) \Bigg]\;.
\end{align}
Detailed derivation and the definition of the terms in Eq.~(\ref{eq.1}) can be seen in Ref.~\cite{ayik2018}.

\subsection{Derivatives of drift coefficients}\label{sec_2.3}
Derivatives of neutron and proton drift coefficients with respect to neutron and proton numbers
cannot be obtained directly from the TDHF calculations. The proton and neutron drift in
the $N-Z$ plane is determined by the potential energy surface of the di-nuclear system.
As a result of the symmetry energy, drift occurs rapidly in the direction perpendicular
to the mean drift path (the beta stability valley), causing rapid equilibration of the
charge asymmetry, and drift continues slowly along the beta stability valley~\cite{merchant1981,merchant1982}. However, when the charge asymmetries of the colliding ions are very close, rapid
equilibration of the charge asymmetry does not occur. For this case, a neighboring system under the same initial conditions could be used for a drift toward the stability valley.
The potential energy surface dominates the nuclear transfers and is given in two
parabolic forms for a di-nuclear system,
\begin{align}
U(N_{1} ,Z_{1} )=\frac{1}{2} aR_{S}^{2} (N_{1} ,Z_{1} )+\frac{1}{2} bR_{V}^{2} (N_{1} ,Z_{1} )\;.
\end{align} 
Here, $R_S(N_1,Z_1)$ and $R_V(N_1,Z_1)$ represent the perpendicular distances of a fragment with
neutron and proton numbers ($N_1$, $Z_1$) from the isoscalar path and the local equilibrium
state along the isoscalar path, respectively. Because of the sharp increase of
asymmetry energy, we expect the isovector curvature parameter $a$ to be much larger
than the isoscalar curvature parameter $b$. Drift coefficients in the over-damped limit
are linked to the potential energy surface in the $(N-Z)$-plane in terms of Einstein
relations as follows~\cite{randrup1979,merchant1981,merchant1982}

\begin{subequations}
\begin{align}\label{eq3}
 v_{n}(t)  &=-\frac{D_{NN}(t) }{T^{*} } \frac{\partial }{\partial N_{1} } U(N_1,Z_1)\\
 v_{z}(t) &=-\frac{D_{ZZ}(t) }{T^{*} } \frac{\partial }{\partial Z_{1}} U(N_1,Z_1)\;,
\end{align}
\end{subequations}
where $T^*$ represents the effective temperature of the system. By using Einstein
relations, the reduced curvature parameters $\alpha=a/T^*$ and $\beta=b/T^*$ are estimated in terms
of drift and diffusion coefficients. As a result of shell effects and microscopic collision
dynamics in the TDHF description, reduced curvature parameters depend on time. In
the macroscopic transport description, we estimate reduced curvature parameters by
carrying out an averaging over a suitable time interval. 

\subsection{Secondary cross-sections of reaction products
}\label{sec_2.4}
The primary fragments produced in the heavy-ion collisions cool down through three main
mechanisms: the emission of light particles, fission or isotope decay, and gamma radiation. We use a statistical GEMINI++ code for the de-excitation
mechanisms of these primary fragments~\cite{charity2008}. The GEMINI++ code uses the Hauser Feshbach formalism~\cite{hauser1952} for the light particle evaporation process, the Bohr-Wheeler
formalism~\cite{moretto1975}, and the Moretto formalism~\cite{moretto1988} for mass symmetric and asymmetric
fission decays, and finally the approach of Blatt and Weisskopf~\cite{blatt1979} for gamma
radiation.
The total excitation energy of a reaction channel is found from the expression $E^*_{tot}=E_\mathrm{c.m.}-TKE(\ell)+Q(N,Z)$
. Here, the total kinetic energy (TKE) and $Q$-value of the
reaction channel for each initial angular momentum are calculated within TDHF. The
total excitation energy and the orbital angular momentum are distributed to the
reaction products in proportion to their masses.
The statistical code GEMINI++ calculates the probability $W_\ell(N,Z \longrightarrow N^{'}, Z^{'})$ of reaching
the final nucleus ($N^{'}, Z^{'}$)
starting from an excited parent nucleus with the number of
neutrons and protons (N, Z) and the excitation energy
$E^*_\ell(N,Z)$
that decays by a
series of consecutive binary decays until it becomes energetically forbidden. Then the
probability distribution of secondary isotopes can be expressed as

\begin{equation}
P_{\ell}^{sec}(N^{'}, Z^{'})= \sum_{N\geq N^{'}}\sum_{Z\geq Z^{'}}(2 \ell+1) P_{\ell}^{pri}(N, Z) W_\ell(N,Z \longrightarrow N^{'}, Z^{'})\;,
\end{equation}
The total secondary isotope production cross-section covering all projectile-like
(PLF) and target-like (TLF) product pairs of the binary nuclear system according to the
probability distributions is as follows:
\begin{equation}
\sigma^{sec}(N^{'}, Z^{'})=\frac{\pi \hbar^{2}}{2 \mu E_\mathrm{c.m.}} \sum_{\ell_{\min }}^{\ell_{\max }}(2 \ell+1) P_{\ell}^{sec}(N^{'}, Z^{'})\;.
\end{equation}

\section{Results and Discussions}
\begin{figure}[!htb]
  \centering
  \includegraphics[width=8.6cm]{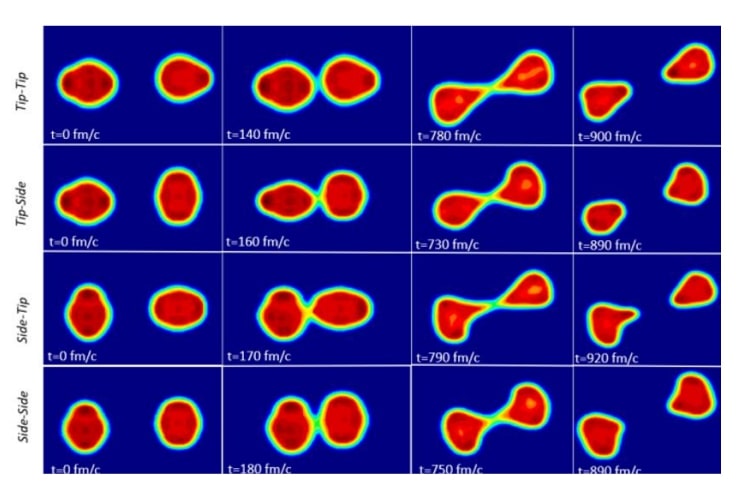}
  \caption{Snapshots of density profiles in the collisions of ${}^{238} \mathrm{U}+{ }^{248} \mathrm{Cm}$ $E_\mathrm{c.m.}=898.7$
MeV) at different times on the reaction plane for $\ell = 100\hbar$.}
  \label{ss}
\end{figure}

\begin{figure}[!htb]
  \centering
  \includegraphics[width=8.6cm]{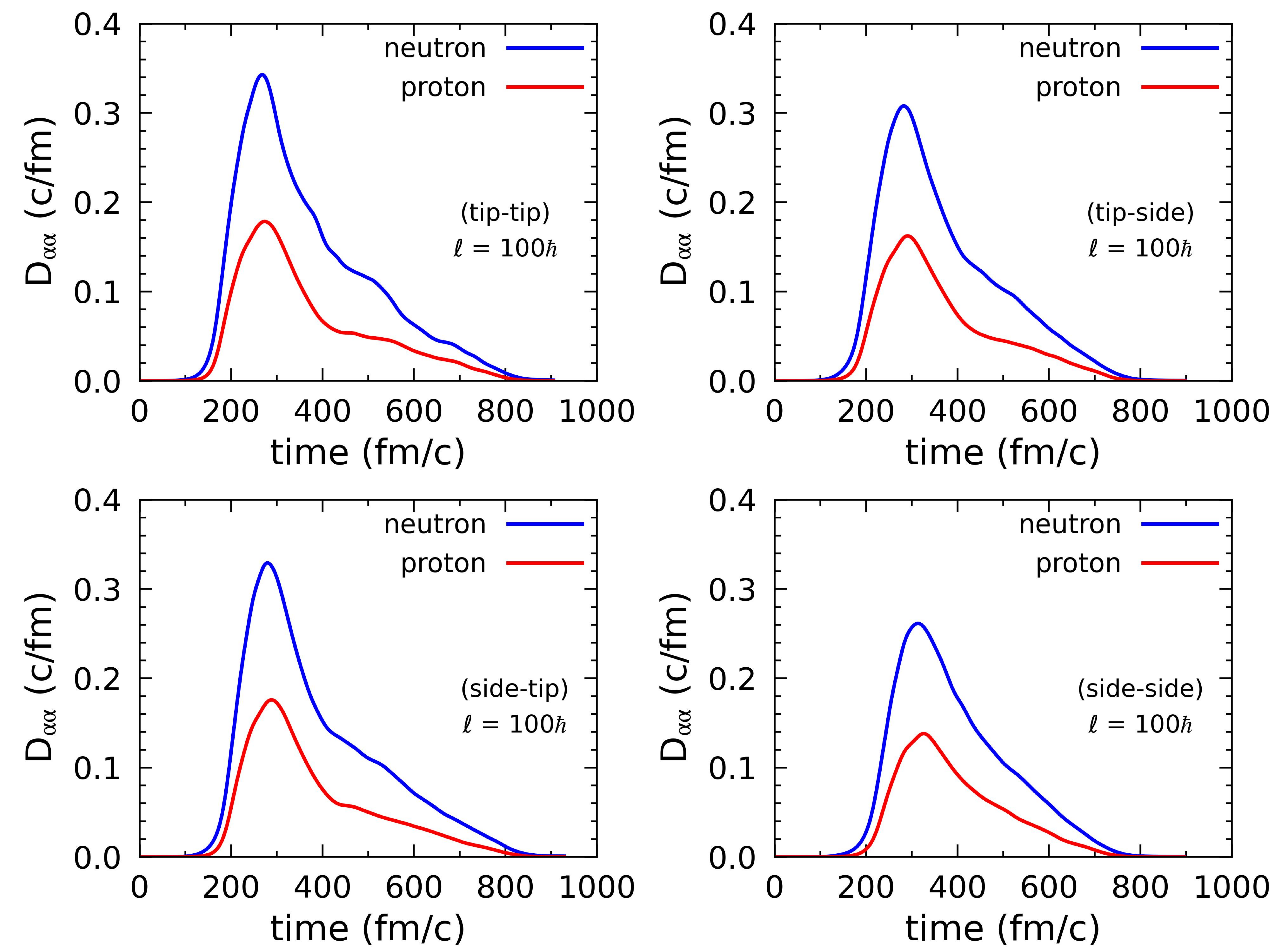}
  \caption{Neutron and proton quantal diffusion coefficient in the ${}^{238} \mathrm{U}+{ }^{248} \mathrm{Cm}$ system at
$E_\mathrm{c.m.}=898.7$~MeV in tip-tip, tip-side, side-tip, and side-side geometries for initial
angular momentum $\ell=100\hbar$.}
  \label{ss_1}
\end{figure}

The mean-field theory describes the one-body dissipation process and the most likely
dynamical path of the collective motion for low-energy heavy-ion collisions. The TDHF theory
is a deterministic approach for many body dynamics since mean-field development
from a given initial condition to a single final state is a deterministic
process. In ${}^{238} \mathrm{U}+{ }^{248} \mathrm{Cm}$, both projectile and target nuclei exhibit strong prolate
deformation in their ground states, and therefore collision dynamics and MNT
mechanism strongly depend on the collision geometry. We examine four different
initial collision configurations at the same energy $ E_\mathrm{c.m.}=898.7$~MeV shown in Fig.~\ref{ss}. We use the letters X and Y to represent the target’s or projectile’s initial orientation,
respectively, perpendicular to and along the beam direction, in analogy with the work
of Kedziora and Simenel of~\cite{kedziora2010}. The symbols XX, XY, YX, and YY stand for four
distinct collision geometries, which correspond to tip-tip, tip-side, side-tip, and side-side geometries, in that order.
The results of TDHF calculations for different values of initial orbital angular
momentum $\ell_i$ are recorded as final values of mass and charge numbers of uranium-like $A_1^f$ and $Z_1^f$ and Curium-like $A_2^f$ and $Z_2^f$ fragments, final total kinetic energy (TKE),
and scattering angles in the center of mass frame (c.m.) for four different collision
geometries. Quantal diffusion coefficients are determined in terms of only occupied
single-particle wave functions of TDHF. As an example, Fig.~\ref{ss_1} shows the neutron and
proton diffusion coefficients of ${}^{238} \mathrm{U}+{ }^{248} \mathrm{Cm}$ system at $E_\mathrm{c.m.}=898.7$~MeV for the initial
angular momentum $\ell=100 \hbar$.

\begin{figure}[!htb]
  \centering
  \includegraphics[width=8.6cm]{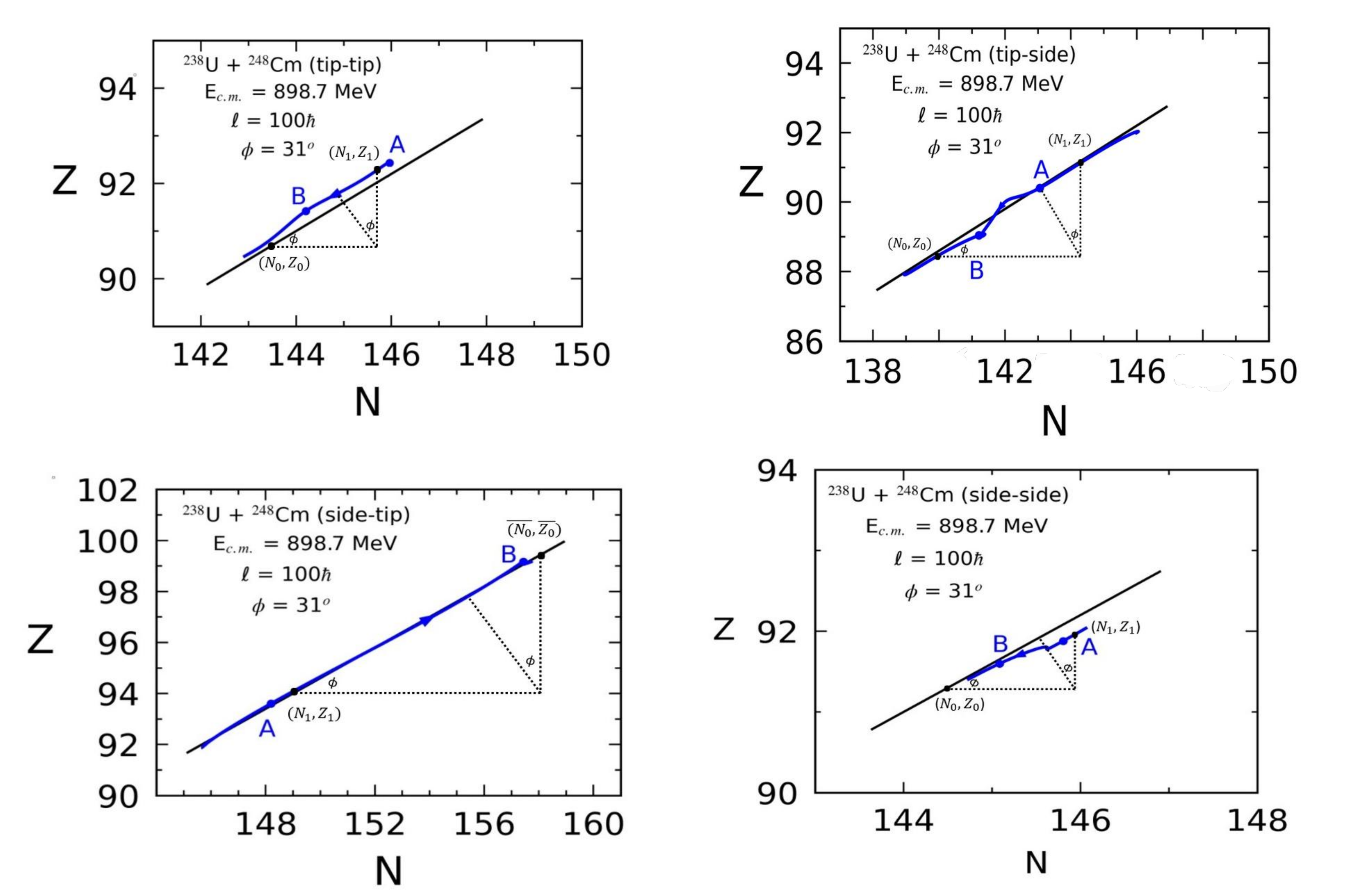}
  \caption{ Mean drift path in $N-Z$ plane for the projectile-like fragments in ${}^{238} \mathrm{U}+{ }^{248} \mathrm{Cm}$
the system at $E_\mathrm{c.m.}=898.7$~MeV in tip-tip, tip-side, side-tip, and side-side geometries.
Solid blue lines denote the mean drift path of U-like fragments, and solid black lines
denote the iso-scalar line.}
  \label{ss_2}
\end{figure}

After colliding nuclei form a dinuclear system, the nucleon drift mechanism is
determined via the potential energy surface in the $(N, Z)$ plane. The drift paths of U-like fragments for each of the four distinct geometries are plotted for the ${}^{238} \mathrm{U}+{ }^{248} \mathrm{Cm}$
system in Fig.~\ref{ss_2} for initial angular momentum $\ell=100\hbar$. In four geometries, that we
consider, the di-nuclear system drifts along the isoscalar path toward a local
equilibrium state. For collisions of actinide nuclei, the lighter local equilibrium state is
located in the vicinity of the lead valley with neutron and proton numbers around $N_0 =
130$, $Z_0 = 82$, and the heavier local equilibrium state is located in the vicinity of
the superheavy valley with neutron and proton numbers around $N_0$ = $N_T - N_0$ = 170, $Z_0$ = $Z_T - Z_0$ = 106. In this system, the initial charge asymmetries ((N-Z)/(N+Z)) of
${}^{238} \mathrm{U}$ is 0.23 and that of ${ }^{248} \mathrm{Cm}$ is 0.23. These nuclei are located on the iso-scalar path with a charge asymmetry of 0.23. As illustrated in Fig.~\ref{ss_2}, the di-nuclear system
formed in the collision of ${}^{238} \mathrm{U}+{ }^{248} \mathrm{Cm}$, U drifts nearly along the iso-scalar path, which
is parallel to the equilibrium valley of stable nuclei. Drift paths contain additional
specific details regarding the nucleon transfer mechanism by illustrating the system’s
evolution in the $N-Z$ plane. The isoscalar path produces an approximately $\phi=31^\circ$
angle with the horizontal neutron axis as it extends completely toward the lead
valley on one end and the superheavy valley on the other.
\begin{figure}[!htb]
  \centering
  \includegraphics[width=8.6cm]{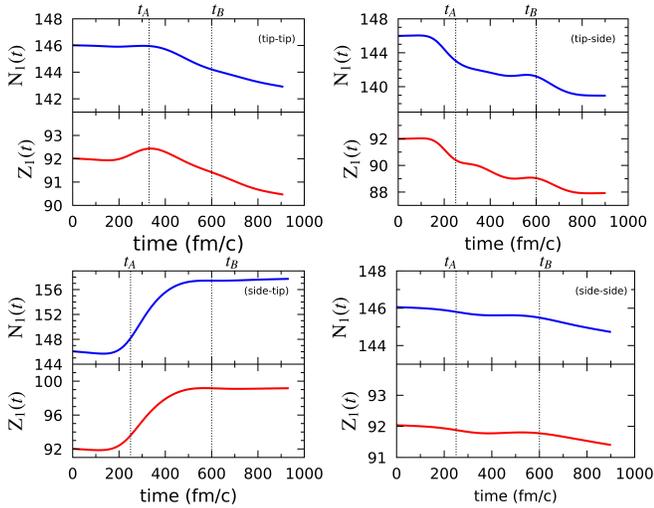}
  \caption{ Mean values of neutron and proton numbers of U -like fragments in the collision
of the ${}^{238} \mathrm{U}+{ }^{248} \mathrm{Cm}$ system at $E_\mathrm{c.m.}=898.7$~MeV in tip-tip, tip-side, side-tip, and side-side collision geometries.}
  \label{ss_3}
\end{figure}
In the ${}^{238} \mathrm{U}+{ }^{248} \mathrm{Cm}$ system, mean values of neutron and proton numbers of projectile-like fragments as a function of time are shown in Fig.~\ref{ss_3} for the initial angular momentum
$\ell=100\hbar$. The labels A and B given in Fig.~\ref{ss_2} and Fig.~\ref{ss_3} indicate the projection of the time
intervals used to determine the isoscalar curvature parameters.

The iso-scalar curvature parameter $\beta$ is calculated by selecting the time intervals $t_A$
and $t_B$ shown in the graph of the average values of the neutron and proton numbers in Fig.~\ref{ss_3},
which correspond to the points $A$ and $B$ on the iso-scalar path given in Fig.~\ref{ss_2}, and the
results are given in Table I.
\begin{figure}[!htb]
  \centering
  \includegraphics[width=8.6cm]{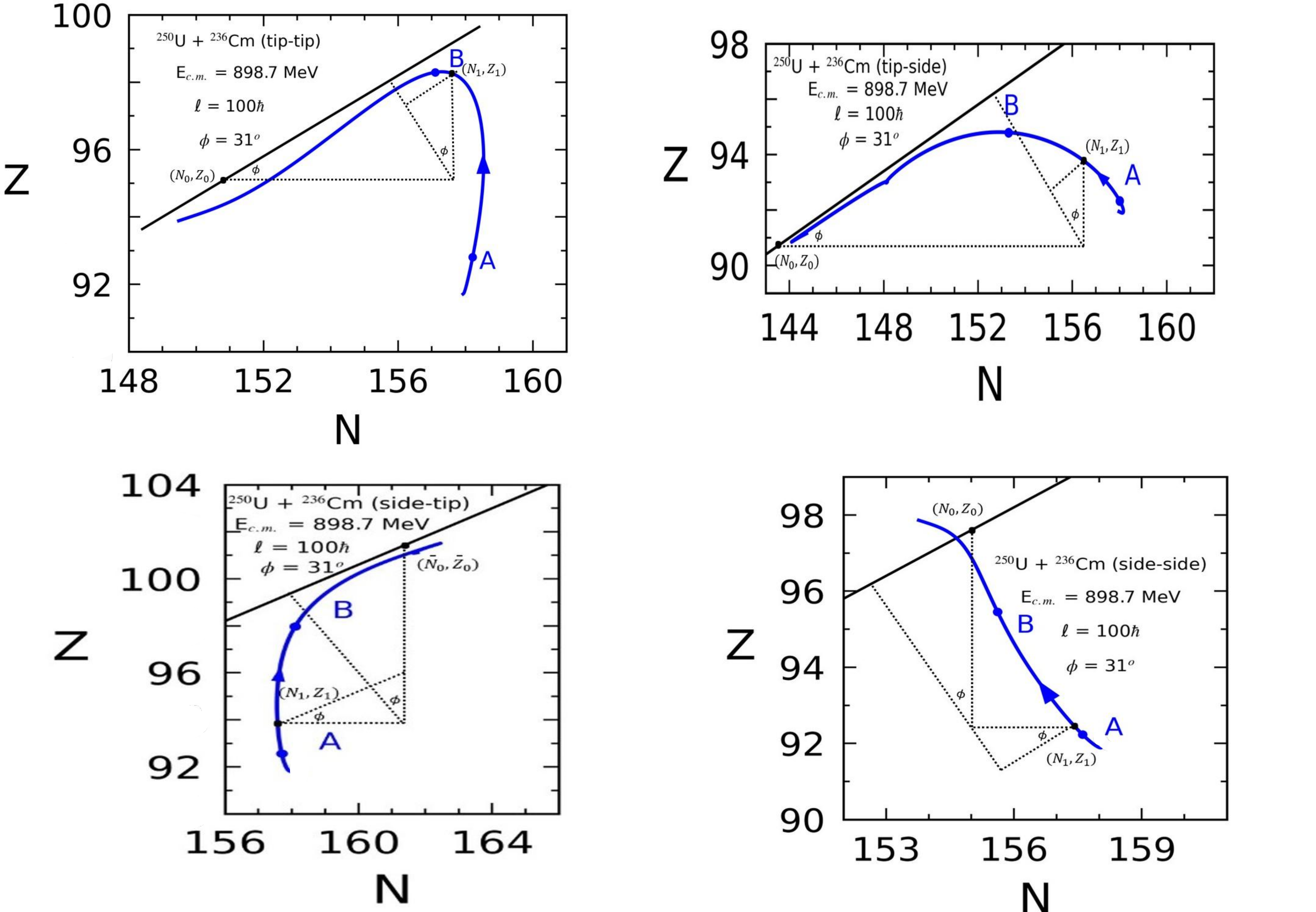}
  \caption{ Mean drift path in N-Z plane for the projectile-like fragments in ${}^{250} \mathrm{U}+{ }^{236} \mathrm{Cm}$
the system at $E_\mathrm{c.m.}=898.7$~MeV in tip-tip, tip-side, side-tip, and side-side geometries.
Solid blue curves denote the mean drift path and black lines denote the iso-scalar line
of U-like fragments. The labels A and B indicate the projection of the time intervals
used to determine the isovector curvature parameters.}
  \label{ss_4}
\end{figure}
\begin{figure}[!htb]
  \centering
  \includegraphics[width=8.6cm]{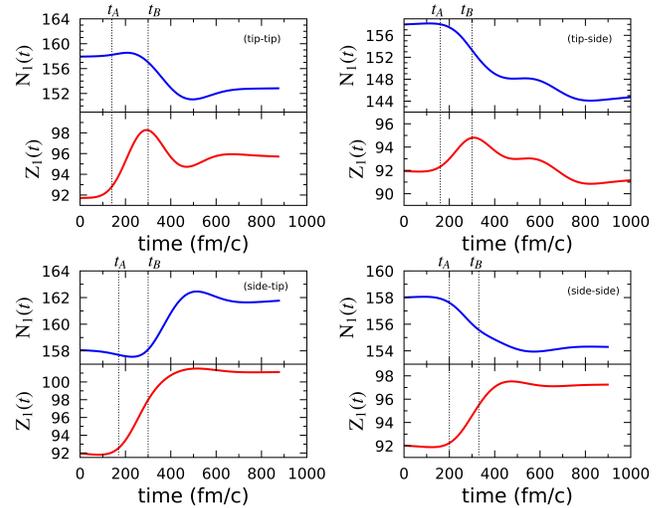}
  \caption{ Mean values of neutron and proton numbers of U -like fragments in the collision
of the ${}^{250} \mathrm{U}+{ }^{236} \mathrm{Cm}$ system at $E_\mathrm{c.m.}=898.7$~MeV in tip-tip, tip-side, side-tip and side-side collision geometries.}
  \label{ss_5}
\end{figure}

To extract information about potential energy in a perpendicular direction to the
stability valley, we need to choose a reaction of a suitable system. For this purpose,
the collision of a neighbor ${}^{250} \mathrm{U}+{ }^{236} \mathrm{Cm}$ system provides such a system. Figure~\ref{ss_4}
shows the drift path of U-like fragments in the collision of ${}^{250} \mathrm{U}+{ }^{236} \mathrm{Cm}$ reaction at tip-tip, tip-side, side-tip, and side-side geometries at $E_\mathrm{c.m.}=898.7$~MeV. The
perpendicular component of this drift line is referred to as the iso-vector path. Then,
the system continues to drift along the iso-scalar path toward symmetry or asymmetry
with the same charge asymmetry and the same slope angle as the iso-scalar path in
the ${}^{238} \mathrm{U}+{ }^{248} \mathrm{Cm}$ system. Before the system reaches local equilibrium, it separates.
By merging the drift data of these two very comparable systems, we can offer a rough
description of the di-nuclear system’s potential energy surface concerning the
equilibrium value. In the ${}^{250} \mathrm{U}+{ }^{236} \mathrm{Cm}$ system, mean values of neutron and proton
numbers of projectile-like fragments as a function of time are shown in Fig.~\ref{ss_5} for initial
angular momentum $\ell=100 \hbar$. The labels $A$ and $B$ given in Fig.~\ref{ss_4} and Fig.~\ref{ss_5} indicate the
projection of the time intervals used to determine the isovector curvature parameters.

When the system drifts toward symmetry on the drift paths, the isoscalar distance from
the local equilibrium state $(N_0,Z_0)$ becomes 
$
R_v(t)=(N_1-N_0)\text{cos}\phi +(Z_1-Z_0)\text{sin}\phi
$ and the isovector distance from the isoscalar path as $
R_s(t)= (Z_1-Z_0)\text{cos}\phi-(N_1-N_0)\text{sin}\phi
$. When drift occurs toward asymmetry, the isoscalar distance from a local equilibrium
state $(\overline{N}_0,\overline{Z}_0)$ becomes $
R_v(t)=(\overline{N}_0-N_1)\text{cos}\phi +(\overline{Z}_0-Z_1)\text{sin}\phi
$ and the isovector
distance as  $
R_s(t)=(\overline{N}_0-N_1)\text{sin}\phi -(\overline{Z}_0-Z_1)\text{cos}\phi
$. The average values of the reduced
curvature parameters are determined by
\begin{align}
\alpha = \frac{1}{t_B-t_A} \int _{t_A}^{t_B}\left(\frac{v_{n} (\tau)\sin \phi }{D_{NN} (\tau)} -\frac{v_{p} (\tau)\cos \phi }{D_{ZZ} (\tau)} \right) \frac {1} {R_{s}} d\tau 
\end{align}
\begin{align}
\beta = \frac{1}{t_A-t_B} \int _{t_A}^{t_B}\left(\frac{v_{n} (\tau)\cos \phi }{D_{NN} (\tau)} +\frac{v_{p} (\tau)\sin \phi }{D_{ZZ} (\tau)} \right) \frac {1} {R_{v}} d\tau \;.
\end{align}

Transport coefficients are time-dependent because of the microscopic shell structure;
hence, reduced curvature parameters are also time-dependent. We utilize constant
curvature parameters and ignore the time dependency in the simple parabolic
parametrization of the potential energy surface. The iso-vector curvature parameter
$\alpha$ is calculated by selecting the time intervals $t_A$ and $t_B$ in the graph of the average
values of the neutron and proton numbers shown in Fig.~\ref{ss_5}, which correspond to the points $A$
and $B$ on the iso-vector path given in Fig.~\ref{ss_4}, as well as the results given in Table I.
\begin{table}[h!]
\label{table1}
\centering

\caption{The calculated curvature parameters for ${}^{238} \mathrm{U}+{ }^{248} \mathrm{Cm}$ and ${}^{250} \mathrm{U}+{ }^{236} \mathrm{Cm}$ systems.}
\begin{tabular}{|c|c|c|c|c|c|c|c|c|}
\hline
 & \multicolumn{4}{c|}{$^{238}\text{U} + ^{248}\text{Cm}$} & \multicolumn{4}{c|}{$^{250}\text{U} + ^{236}\text{Cm}$} \\ \hline
\textbf{Orientation} & XX & XY & YX & YY & XX & XY & YX & YY \\ \hline
$t_A$ (fm/c) & \rule{0pt}{2.5ex} 330 & 250 & 250 & 250 & 140 & 160 & 170 & 250 \\ \hline
$t_B$ (fm/c) & 600 & 600 & 600 & 750 & 300 & 300 & 300 & 350 \\ \hline
$\beta$ & 0.004 & 0.004 & 0.010 & 0.004 &  &  &  &  \\ \hline
$\alpha$ &  &  &  &  & 0.177 & 0.108 & 0.177 & 0.120 \\ \hline
  
\end{tabular}
\end{table}

\begin{figure}[!htb]
  \centering
  \includegraphics[width=8.6cm]{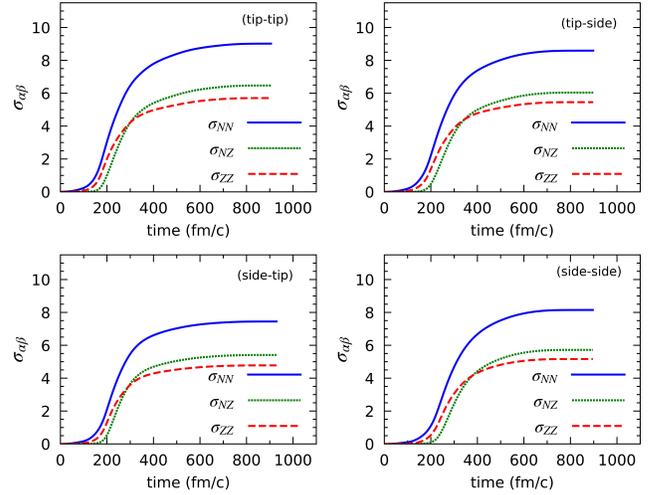}
  \caption{Neutron, proton, and mixed variances as a function of time in the ${}^{238} \mathrm{U}+{ }^{248} \mathrm{Cm}$,
 $\ell=100 \hbar$. the system at $E_\mathrm{c.m.}=898.7$~MeV in tip-tip, tip-side, side-tip, and side-side collision
geometries.}
  \label{ss_6}
\end{figure}

Variances and covariances are determined from the solutions of the coupled
differential equations  given in Eqs. (\ref{Eq:Coupled1},\ref{Eq:Coupled2},\ref{Eq:Coupled3}) with initial conditions $\sigma_{NN}^2$ (t = 0) = 0, $\sigma_{ZZ}^2$ (t =0) = 0, and $\sigma_{NZ}^2$
(t = 0) = 0, for each orbital angular momentum. As an example, Fig.~\ref{ss_6} shows neutron,
proton, and mixed variances, as a function of time, in the collision of ${}^{238} \mathrm{U}+{ }^{248} \mathrm{Cm}$
system at $E_\mathrm{c.m.}=898.7$~MeV for the tip-tip tip-side, side-tip, and side-side collision
geometries for $\ell=100 \hbar$. In the initial phase of the reaction, up to about $t=300$~fm/c, we
see that the magnitudes of the variances are in the order $\sigma_{NZ}<\sigma_{ZZ}<\sigma_{NN}$. The
correlation develops over time and the order changes to $\sigma_{ZZ}<\sigma_{NZ}<\sigma_{NN}$. This
shows the importance of correlations after significant energy dissipation. The neutron-proton $\sigma_{NZ}$ correlation is zero in standard mean-field theories, and the effect of
fluctuations does not agree with the observation in dissipative systems. In the
stochastic mean-field approach, the $\sigma_{NZ}$ correlation is not zero, and the effect of
fluctuations agrees with the observation.

We present the numerical results of TDHF and SMF calculations for ${}^{238} \mathrm{U}+{ }^{248} \mathrm{Cm}$
reaction at $E_\mathrm{c.m.}=898.7$~MeV in Table~\ref{table_2} and Table~\ref{table_3}. The tables include asymptotic
values of neutron dispersion $\sigma_{NN}(\ell)$, proton dispersion $\sigma_{ZZ}(\ell)$, mixed dispersions
$\sigma_{NZ}(\ell)$,  for each orbital
angular momentum. The initial orbital angular momentum has been selected in
increments of 20$\hbar$ and then 40$\hbar$. The TDHF code~\cite{umar1991a,umar2006c} was used in the computations using the SLy4d Skyrme energy density functional
\cite{kim1997}, with a box size of 60 × 60 × 36~fm in the x-y-z directions, respectively.
\begin{table}[!t]
\caption{Results of the TDHF and SMF calculations for the ${}^{238}\text{U}+{}^{248}\text{Cm}$ system at $E_\mathrm{c.m.}=898.7$~MeV in tip-tip (XX), tip-side (XY) geometries.}
\label{table_2}
\setlength{\tabcolsep}{3pt} 
\renewcommand{\arraystretch}{1.2} 
\centering
\begin{ruledtabular}
\begin{tabular}{c | c c c c c c c c c c}
- & $\ell_i$ $(\hbar)$ & $Z_1^f$ & $A_1^f$ & $Z_2^f$ & $A_2^f$ & $TKE$ & $\sigma_{NN}$ & $\sigma_{ZZ}$ & $\sigma_{NZ}$ \\
\hline
\multirow{10}{*}{\textbf{XX}} 
& 0   & 93.3 & 240.8 & 94.7 & 245.2 & 651.6 & 8.7 & 5.5 & 6.2 \\
& 20  & 93.2 & 240.5 & 94.8 & 245.2 & 623.9 & 8.8 & 5.5 & 6.2 \\
& 40  & 92.7 & 239.4 & 95.2 & 246.6 & 628.4 & 8.8 & 5.5 & 6.2 \\
& 60  & 92.1 & 237.5 & 95.9 & 248.5 & 630.9 & 8.8 & 5.5 & 6.3 \\
& 80  & 91.3 & 235.5 & 96.6 & 250.4 & 619.5 & 8.9 & 5.6 & 6.3\\
& 100 & 90.6 & 233.6 & 97.4 & 252.4 & 592.9 & 9.0 & 5.7 & 6.5\\
& 120 & 90.2 & 232.4 & 97.8 & 253.6 & 565.7 & 9.1 & 5.7 & 6.5\\
& 140 & 90.4 & 233.3 & 97.6 & 252.7 & 545.0 & 9.2 & 5.8 & 6.6\\
& 160 & 90.4 & 233.3 & 97.6 & 252.7 & 536.8 & 9.2 & 5.8 & 6.6\\
& 180 & 90.2 & 232.5 & 97.8 & 253.5 & 543.4 & 9.1 & 5.8 & 6.6\\
&200 &90.7 &234.0 &97.3 &252.0 &557.9 &9.1 &5.7 &6.5\\
&240 &91.8 &236.9 &96.1 &249.1 &573.1 &8.9 &5.6 &6.3\\
&280 &92.5 &238.5 &95.5 &247.5 &592.7 &8.5 &5.4 &6.0\\
&320 &92.4 &237.8 &95.6 &248.2 &637.5 &8.0 &5.0 &5.5\\
&360 &92.0 &237.1 &96.0 &257.1 &687.0 &7.1 &4.5 &4.8\\
&400 &91.2 &235.3 &96.8 &250.7 &737.1 &6.0 &3.8 &3.7\\
&440 &91.5 &236.3 &96.5 &249.7 &806.7 &4.5 &2.8 &2.1\\
&480 &91.9 &237.7 &96.1 &248.3 &876.6 &2.5 &1.3 &0.5\\
&520 &92.0 &237.9 &96.0 &248.1 &889.4 &1.6 &0.8 &0.2\\
&560 &92.0 &238.0 &96.0 &248.0 &891.1 &1.2 &0.5 &0.1\\
&600 &92.0 &238.0 &96.0 &248.0 &892.1 &1.1 &0.4 &0.1\\
\hline
\multirow{10}{*}{\textbf{XY}} 
& 0   & 86.9 & 223.6 & 101.1 & 262.4 & 653.9 & 8.5 & 5.4 & 6.0\\
& 20  & 86.9 & 223.8 & 101.0 & 262.2 & 655.4 & 8.5 & 5.4 & 6.0\\
& 40  & 87.1 & 224.3 & 101.0 & 261.7 & 654.2 & 8.5 & 5.4 & 6.0\\
& 60  & 87.3 & 225.0 & 100.7 & 261.0 & 651.6 & 8.5 & 5.4 & 6.0\\
& 80  & 87.6 & 225.4 & 100.4 & 260.2 & 646.2 & 8.5 & 5.4 & 5.9\\
& 100 & 87.9 & 226.6 & 100.1 & 259.4 & 638.7 & 8.6 & 5.5 & 6.0\\
& 120 & 88.2 & 227.6 & 99.7 & 258.4 & 626.0 & 8.6 & 5.5 & 6.1\\
& 140 & 88.5 & 228.5 & 99.5 & 257.5 & 611.3 & 8.6 & 5.5 & 6.1\\
& 160 & 89.0 & 229.7 & 99.0 & 256.3 & 598.0 & 8.7 & 5.5 & 6.1\\
& 180 & 89.7 & 231.6 & 98.3 & 254.4 & 588.8 & 8.6 & 5.5 & 6.1\\
&200 &90.6 &234.2 &97.4 &251.8 &585.8 &8.6 &5.4 &6.2\\
&240 &91.8 &237.5 &96.2 &248.5 &602.5 &8.4 &5.3 &5.8\\
&280 &91.5 &236.0 &96.5 &250.0 &620.5 &8.1 &5.2 &5.6\\
&320 &91.9 &237.0 &96.1 &249.0 &646.4 &7.7 &4.9 &5.2\\
&360 &92.7 &239.0 &95.3 &284.7 &674.4 &7.2 &4.6 &4.8\\
&400 &92.9 &239.4 &95.0 &246.6 &703.2 &6.5 &4.2 &4.2\\
&440 &93.0 &239.6 &95.0 &246.4 &740.3 &5.8 &3.7 &3.4\\
&480 &92.9 &239.6 &95.1 &246.4 &782.9 &4.9 &3.1 &2.4\\
&520 &92.5 &238.9 &95.5 &247.0 &829.0 &3.8 &2.3 &1.4\\
&560 &92.1 &238.2 &95.9 &247.8 &870.9 &2.3 &1.2 &0.4\\
&600 &92.0 &238.0 &96.0 &248.0 &886.6 &1.6 &0.8 &0.2\\
\end{tabular}
\end{ruledtabular}
\end{table}

\begin{table}[!t]
\caption{ Results of the TDHF and SMF calculations for the ${}^{238}\text{U}+{}^{248}\text{Cm}$ system at $E_\mathrm{c.m.}=898.7$~MeV in side-tip (YX) and side-side (YY) geometries.}
\label{table_3}
\setlength{\tabcolsep}{3pt} 
\renewcommand{\arraystretch}{1.2} 
\centering
\begin{ruledtabular}
\begin{tabular}{c | c c c c c c c c c c}
- & $\ell_i$ $(\hbar)$ & $Z_1^f$ & $A_1^f$ & $Z_2^f$ & $A_2^f$ & $TKE$ & $\sigma_{NN}$ & $\sigma_{ZZ}$ & $\sigma_{NZ}$ \\
\hline
\multirow{10}{*}{\textbf{YX}} 
&0 &88.0 &226.3 &100.0 &259.6 &627.7 &7.3 &4.7 &5.3\\
&20 &88.0 &226.3 &100.0 &259.6 &627.9 &7.3 &4.7 &5.3\\
&40 &87.9 &226.2 &100.0 &259.8 &625.3 &7.4 &4.7 &5.3\\
&60 &87.9 &226.5 &100.0 &259.4 &620.4 &7.4 &4.7 &5.3\\
&80 &88.1 &227.3 &99.9 &258.7 &610.3 &7.4 &4.8 &5.4\\
&100 &88.9 &229.2 &99.1 &256.8 &594.5 &7.5 &4.8 &5.4\\
&120 &90.2 &232.9 &97.8 &253.1 &582.4 &7.5 &4.8 &5.4\\
&140 &91.6 &236.9 &96.4 &249.1 &581.6 &7.5 &4.8 &5.4\\
&160 &92.4 &239.2 &95.6 &246.8 &588.0 &7.5 &4.8 &5.4\\
&180 &92.9 &240.7 &95.1 &245.3 &595.3 &7.4 &4.8 &5.4\\
&200 &93.3 &241.8 &94.7 &244.2 &604.3 &7.4 &4.7 &5.3\\
&240 &93.5 &241.8 &94.5 &244.2 &611.8 &7.2 &4.6 &5.2\\
&280 &93.8 &242.8 &94.2 &243.2 &615.1 &7.0 &4.5 &5.1\\
&320 &92.7 &240.0 &95.3 &246.0 &642.4 &6.7 &4.3 &4.8\\
&360 &92.2 &238.4 &95.8 &247.6 &680.9 &6.3 &4.1 &4.4\\
&400 &91.8 &237.3 &96.2 &248.7 &715.7 &5.8 &3.7 &4.0\\
&440 &91.3 &236.0 &96.7 &249.9 &749.2 &5.2 &3.4 &3.3\\
&480 &91.2 &235.6 &96.8 &250.4 &783.6 &4.4 &2.9 &2.6\\
&520 &91.3 &235.9 &96.7 &250.0 &824.2 &3.6 &2.3 &1.7\\
&560 &91.6 &236.8 &96.4 &249.2 &857.4 &2.7 &1.6 &0.9\\
&600 &91.9 &237.7 &96.1 &248.0 &882.3 &1.7 &0.8 &0.3\\
\hline
\multirow{10}{*}{\textbf{YY}} 
&0 &91.6 &236.3 &96.4 &249.6  &667.5 &8.1 &5.1 &5.7\\
&20 &91.6 &236.4 &96.3 &249.6 &668.7 &8.1 &5.1 &5.7\\
&40 &91.6 &236.5 &96.4 &249.5 &668.5 &8.1 &5.2 &5.7\\
&60 &91.6 &236.6 &96.4 &249.4 &668.6 &8.2 &5.2 &5.7\\
&80 &91.6 &236.6 &96.4 &249.4 &666.2 &8.2 &5.2 &5.7\\
&100 &90.6 &233.6 &97.4 &252.4 &665.3 &8.2 &5.2 &5.7\\
&120 &90.2 &236.4 &96.6 &249.6 &660.7 &8.1 &5.2 &5.7\\
&140 &91.5 &236.6 &96.5 &249.4 &656.7 &8.1 &5.1 &5.7\\
&160 &91.7 &237.1 &96.3 &248.9 &653.7 &8.1 &5.1 &5.6\\
&180 &92.1 &237.9 &95.9 &248.1 &650.3 &8.0 &5.1 &5.6\\
&200 &92.5 &238.9 &95.5 &247.1 &648.9 &7.9 &5.0 &5.5\\
&240 &93.0 &240.3 &95.0 &245.7 &651.2 &7.7 &4.9 &5.3\\
&280 &92.7 &239.9 &95.5 &246.1 &661.6 &7.3 &4.7 &4.9\\
&320 &92.6 &239.5 &95.4 &246.5 &670.0 &6.9 &4.4 &4.6\\
&360 &95.5 &239.1 &95.5 &246.9 &691.0 &6.4 &4.1 &4.1\\
&400 &92.3 &238.6 &95.7 &247.4 &734.9 &5.8 &3.7 &3.5\\
&440 &92.2 &238.1 &95.8 &247.9 &772.7 &5.1 &3.2 &2.8\\
&480 &92.1 &237.7 &95.9 &248.3 &810.7 &4.2 &2.6 &1.9\\
&520 &91.9 &237.5 &96.1 &248.5 &846.1 &3.8 &1.9 &1.1\\
&560 &91.9 &237.6 &96.1 &248.4 &871.9 &2.5 &1.3 &0.5\\
&600 &92.0 &237.8 &96.0 &248.2 &882.7 &1.9 &1.0 &0.3\\
\end{tabular}
\end{ruledtabular}
\end{table}

\begin{figure}[!htb]
  \centering
  \includegraphics[width=8.6cm]{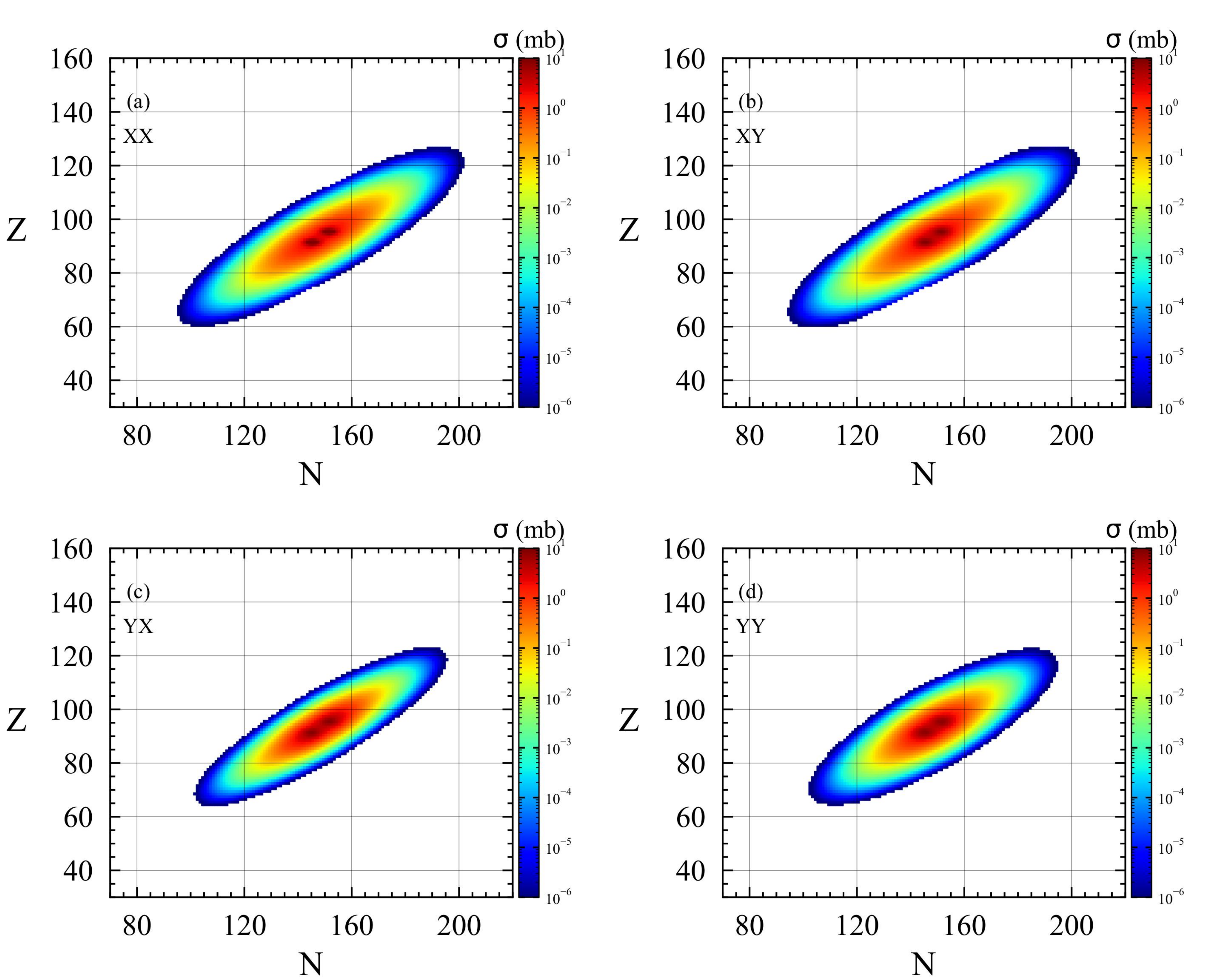}
  \caption{Primary production cross sections in the N-Z plane for the ${}^{238} \mathrm{U}+{ }^{248} \mathrm{Cm}$ system
at $E_\mathrm{c.m.}=898.7$~MeV in tip-tip (XX), tip-side (XY), side-tip (YX), and side-side (YY)
geometries}
  \label{ss_7}
\end{figure}
\begin{figure}[!htb]
  \centering
  \includegraphics[width=8.6cm]{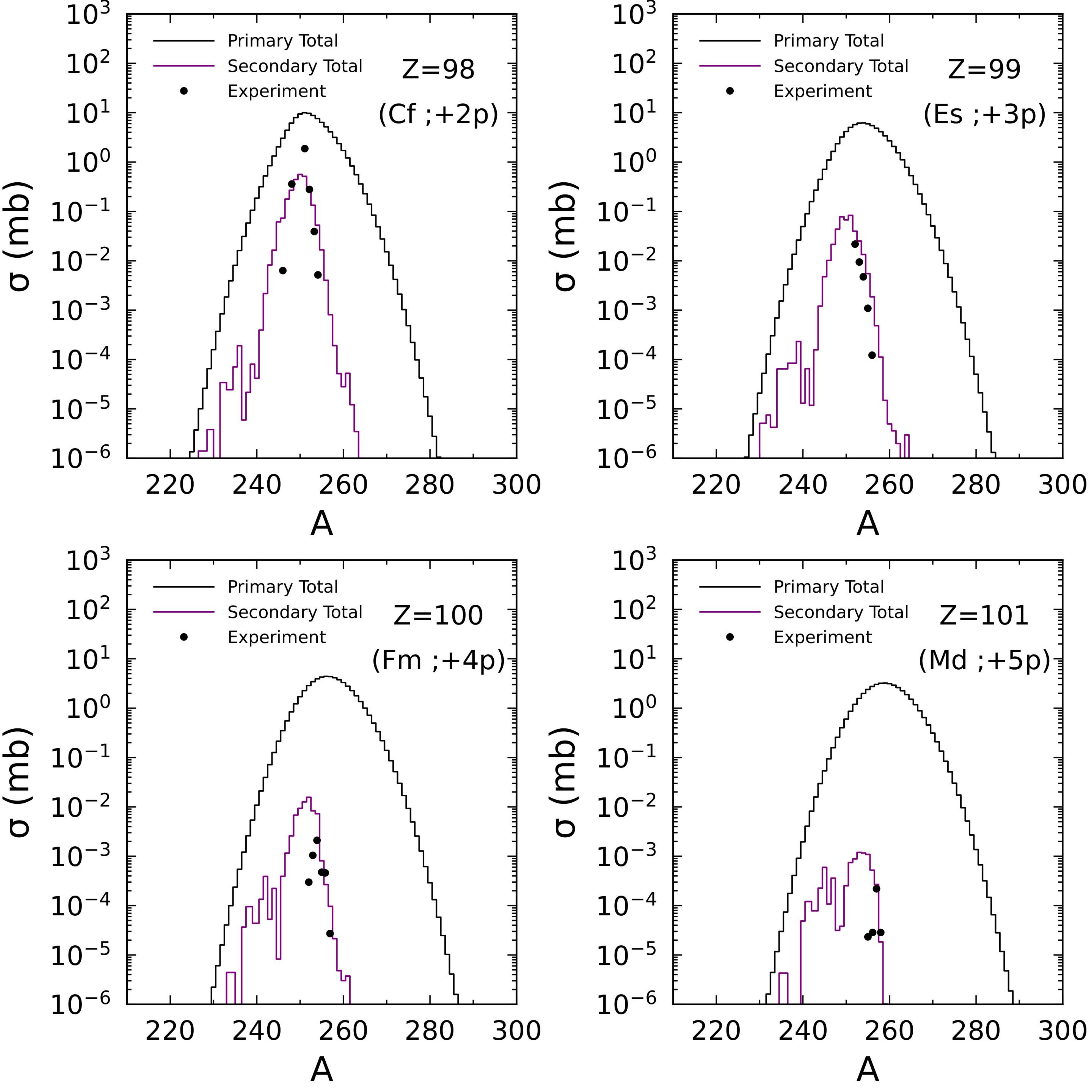}
  \caption{Production cross sections of transuranium isotopes in the ${}^{238} \mathrm{U}+{ }^{248} \mathrm{Cm}$ 
system at $E_\mathrm{c.m.}=898.7$~MeV in tip-tip, tip-side, side-tip, and side-side collision
geometries.}
  \label{ss_8}
\end{figure}

In calculations, we perform a summation in the range from $\ell_{min}=0\hbar$ to $\ell_{max}=600\hbar$.
We calculate total double cross-sections for tip-tip, tip-side, side-tip, and side-side
geometries. Fig.~\ref{ss_7} shows double cross-sections in $(N-Z)$ plane for the production of
primary fragments in tip-tip (a), tip-side (b), side-tip (c), and side-side (d) geometries.
Large values of mixed variances indicate strong correlations in neutron-proton
transfers. As a result, the major axes of equal cross-section elliptic curves are aligned
along the valley of stability.
The total cross-section calculations of secondary fragments are given in Fig.~\ref{ss_8},
assuming equal weights for contributions from each initial geometry. The comparison
of the calculated critical values with the experimental values is given in Table.~\ref{table_4}. As the
number of transferred nucleons increases, the cross-section decreases. The SMF
results are consistent with the experimental values in terms of magnitude and peak
points. The SMF results were successful in explaining the isotope cross-sections
measured in the experiment.
\begin{table}[h!]
\centering
\label{table_4}
\caption{Isotope Peak Points and Measured Values}
\begin{minipage}{0.6\textwidth} 
\raggedright 
\small 
\setlength{\tabcolsep}{3pt} 
\renewcommand{\arraystretch}{1.1} 
\begin{tabular}{|c|c|c|c|c|c|c|}
\hline
 & \multicolumn{2}{c|}{Primary Isotope} & \multicolumn{2}{c|}{Secondary Isotope} & \multicolumn{2}{c|}{Measured Values} \\
\hline
  Z & Peak, A & $\sigma$ (mb) & Peak, A & $\sigma$ (mb) & Peak, A & $\sigma$ (mb) \\
\hline
98  & 251 & 10.4 & 250 & 0.60  & 251 & 1.87 \\
\hline
99  & 253 & 6.20 & 251 & 0.08  & 252 & 0.02 \\
\hline
100 & 256 & 4.40 & 252 & 0.02  & 254 & 0.002 \\
\hline
101 & 259 & 3.20 & 253 & 0.001 & 257 & 0.0002 \\
\hline
\end{tabular}
\end{minipage}
\end{table}

\begin{figure}[!htb]
  \centering
  \includegraphics[width=8.6cm]{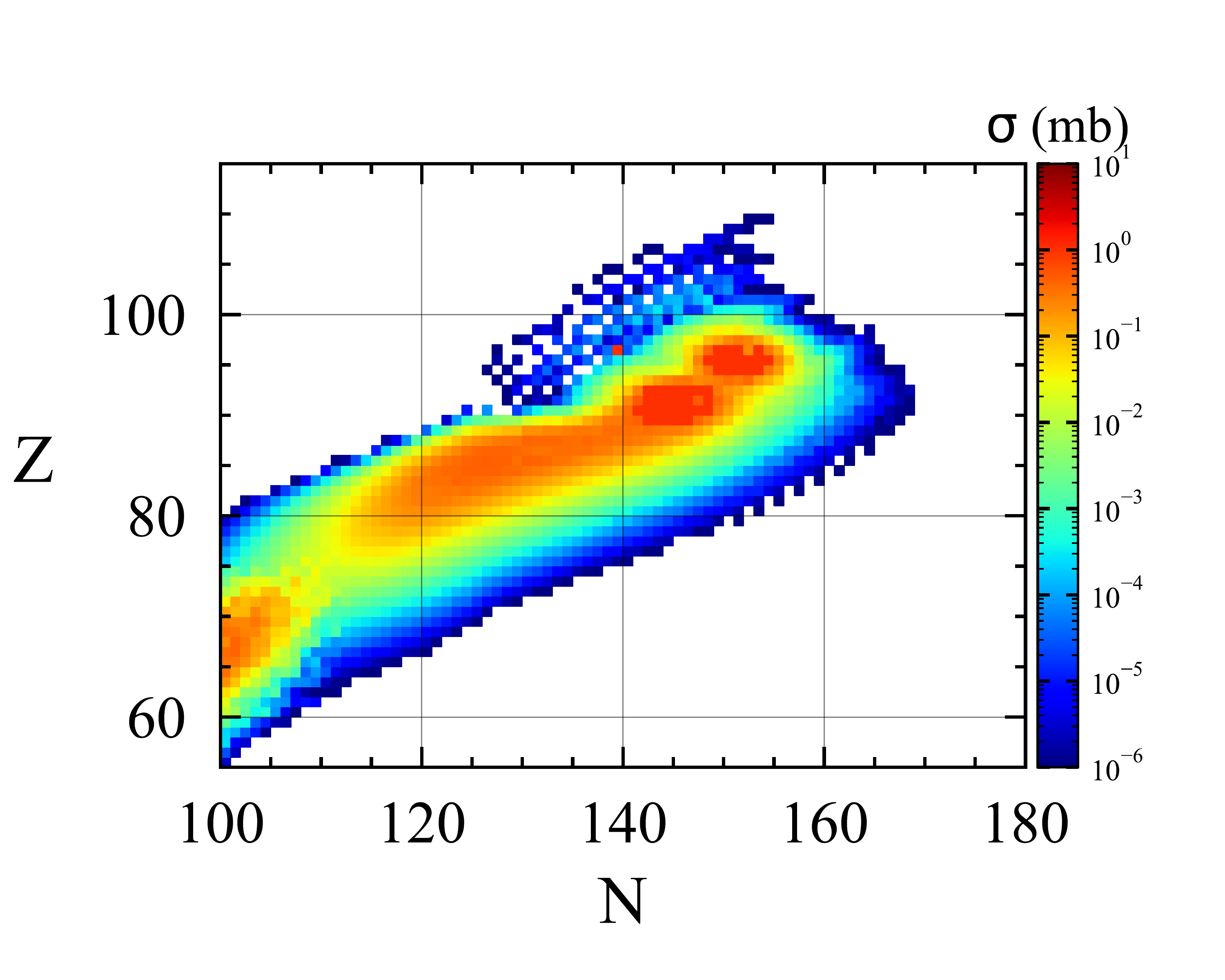}
  \caption{Calculated total production cross sections of secondary isotopes in the (N-Z)
plane for the ${}^{238} \mathrm{U}+{ }^{248} \mathrm{Cm}$  system at $E_\mathrm{c.m.}=898.7$~MeV energy}
  \label{ss_9}
\end{figure}

The production of isotopes in the projectile-target combinations has been predicted
theoretically. The results are obtained by using the quantal diffusion approach based on the
SMF for primary production and secondary calculations by employing the GEMINI++
code. In Fig.~\ref{ss_9}, the calculated cross-sections for the transfer channels (+2p; Cf), (+3p;
Es), (+4p; Fm), (+5p; Md) for the reaction ${}^{238} \mathrm{U}+{ }^{248} \mathrm{Cm}$  are compared with the
experimental data from reference~\cite{kratz2013}. For $Z=98$ (Californium), the primary total data shows a peak
cross-section of 10.4~mb, corresponding to a mass number of 251. This indicates
that, during the initial stages of the reaction, fragment production is most likely around
this mass number, suggesting a preferred nucleon exchange and fragmentation
pathway for Californium isotopes. In contrast, the secondary total data for $Z=98$
reveals a lower cross-section value of 0.60~mb, associated with a mass number of
250, reflecting the redistribution of nucleons and energy during the de-excitation
process. For $Z=99$ (Einsteinium), the primary total data indicates a mass number of
253 with a corresponding cross-section of 6.20~mb, suggesting a significant likelihood
of fragment production for this mass number during the initial reaction stages. For the
secondary total data indicates a mass number of 251 with a corresponding cross
section of 0.08~mb. Turning to $Z=100$ (Fermium), the primary total data reveals a mass
number of 256 with a cross-section of 4.40~mb, indicating a notable probability for
fragment production in this region. The secondary total data for $Z=100$, however,
shows a mass number of 252 and a significantly lower cross-section of 0.02~mb,
highlighting a decrease in fragment production likelihood during the secondary stages.
For $Z=101$ (Mendelevium), the primary total data reflects a mass number of 259 with
a cross-section of 3.20~mb, suggesting a consistent fragmentation pathway in the initial
reaction. Conversely, the secondary total data for Z=101 shows a mass number of
253 with a minimal cross-section of 0.001~mb, emphasizing a marked reduction in
the probability of fragment production in this mass range during the later stages. These
findings underscore the complex behavior of Californium, Einsteinium, Fermium, and
Mendelevium during nuclear reactions, with distinct mass distributions evolving from
the primary to secondary stages of fragmentation. The data illustrate how variations in $Z$ influences the outcomes of nuclear reactions
modeled by SMF+GEMINI. It is observed that the calculated results align closely with
the experimental data, indicating the reliability of the theoretical models used in this
analysis. This agreement is particularly evident in the behavior of the cross sections, which exhibit a rapid decrease as the mass number of the produced fragments
increases. As the atomic number of the surviving heavy nuclei rises, the cross sections
shift toward regions of higher mass numbers. This trend suggests that the increasing
stability and interaction characteristics of heavier nuclei play a significant role in
fragment production.

The calculated total production cross-sections of secondary isotopes in the $(N-Z)$ plane
for the transuranium and superheavy regions for the ${}^{238} \mathrm{U}+{ }^{248} \mathrm{Cm}$ system at
$E_\mathrm{c.m.}=898.7$~MeV energy are given in Fig.~\ref{ss_9}. The isotope production cross sections can be obtained in the millibarn range for the Z=93-98 region, in the microbarn
range in the $Z=99-102$ region, and in the nanobarn range for the $Z=103-110$ region. It
was shown that the ${}^{238} \mathrm{U}+{ }^{248} \mathrm{Cm}$ system at $E_\mathrm{c.m.}=898.7$~MeV energy is a suitable
system for the synthesis of transuranium neutron-rich superheavy isotopes.

\begin{figure}[!htb]
  \centering
  \includegraphics[width=8.6cm]{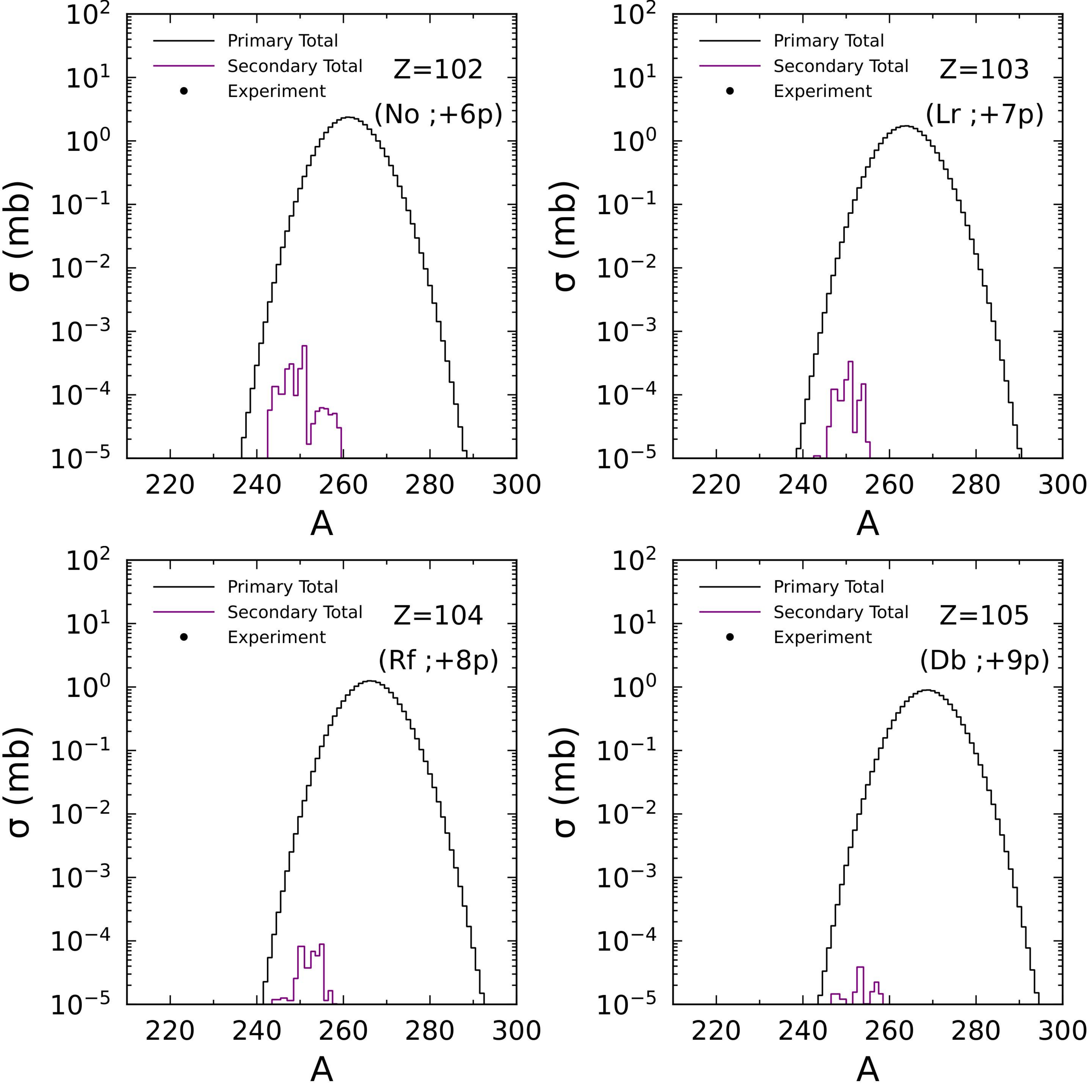}
  \caption{SMF+GEMINI++ calculations of total production cross-sections of $Z=102-105$
isotopes in the ${}^{238} \mathrm{U}+{ }^{248} \mathrm{Cm}$  $E_\mathrm{c.m.}=898.7$ reaction}
  \label{ss_10}
\end{figure}

In low-energy collisions, the cross-section is very sensitive to shell effects. During the
de-excitation processes in the MNT reaction, the cross-section values decrease due to
strong shell effects. In this case, the expected magic number isotopes (Z=114, N=184)
in the island of stability, in the superheavy region, could not be reached. Theoretical
results calculated for the $Z=102-105$ region, for which there are no experimental data,
are given in Fig.~\ref{ss_10}, and it was found that the cross-section values would be smaller
than the microbarn level.

The decay processes of MNT products are related to the local excitation energy and
angular momentum at the equilibrium time, fission barrier, and separation energy of
evaporated particles (neutron, proton, deuterium, triton, alpha, etc.). A strong shell
effect can increase the fission barrier and separation energy, which ensures the
survival of the primary fragment. Shell effect can increase MNT yields by up to two
orders of magnitude. In the absence of shell effect, the production cross sections of
primary and secondary fragments decrease monotonically with mass or charge
number, which reduces the formation of heavy rare isotopes in MNT reactions.

\section{Conclusions}\label{sec_4}
In this study, we present an investigation of the multinucleon transfer mechanism in collisions of ${}^{238} \mathrm{U}+{ }^{248} \mathrm{Cm}$  system at $E_\mathrm{c.m.}=898.7$~MeV employing a quantal diffusion description, which provides a source for developing fluctuations, based on the SMF approach beyond the TDHF theory. The standard mean-field description of TDHF determines the average evolution of the most probable path of heavy-ion collision dynamics, however, the SMF provides an extension of the standard TDHF description by including mean-field fluctuations consistent with the fluctuation-dissipation theorem of non-equilibrium statistical mechanics. Transport coefficients include quantal effects due to shell structure and Pauli blocking, and do not involve any adjustable parameters other than the standard parameters of the effective Skyrme force used in TDHF description.

Highly excited primary fragments decay by particle emission and secondary fission. The de-excitation processes of primary fragments are followed with statistical code GEMINI++ and the production cross-sections of the secondary fragments are calculated. The observed agreement between the calculated and experimental data shows that the SMF approach is a powerful way to explain the MNT process. The SMF approach is an effective microscopic and quantal approach for the synthesis of neutron-rich isotopes farther from the target or superheavy elements by MNT reactions.

\section{Acknowledgements}\label{sec_5}
S.A. gratefully acknowledges Middle East Technical University for the warm hospitality extended to him during his visits. This work is supported in part by US DOE Grants No. DE-SC0013847. This work is supported in part by TUBITAK Grant No. 122F150. The numerical calculations reported in this paper were partially performed at TUBITAK ULAKBIM, High Performance and Grid Computing Center (TRUBA resources).

\bibliography{VU_bibtex_master}

\begin{thebibliography}{64}%
\makeatletter
\providecommand \@ifxundefined [1]{%
 \@ifx{#1\undefined}
}%
\providecommand \@ifnum [1]{%
 \ifnum #1\expandafter \@firstoftwo
 \else \expandafter \@secondoftwo
 \fi
}%
\providecommand \@ifx [1]{%
 \ifx #1\expandafter \@firstoftwo
 \else \expandafter \@secondoftwo
 \fi
}%
\providecommand \natexlab [1]{#1}%
\providecommand \enquote  [1]{``#1''}%
\providecommand \bibnamefont  [1]{#1}%
\providecommand \bibfnamefont [1]{#1}%
\providecommand \citenamefont [1]{#1}%
\providecommand \href@noop [0]{\@secondoftwo}%
\providecommand \href [0]{\begingroup \@sanitize@url \@href}%
\providecommand \@href[1]{\@@startlink{#1}\@@href}%
\providecommand \@@href[1]{\endgroup#1\@@endlink}%
\providecommand \@sanitize@url [0]{\catcode `\\12\catcode `\$12\catcode
  `\&12\catcode `\#12\catcode `\^12\catcode `\_12\catcode `\%12\relax}%
\providecommand \@@startlink[1]{}%
\providecommand \@@endlink[0]{}%
\providecommand \url  [0]{\begingroup\@sanitize@url \@url }%
\providecommand \@url [1]{\endgroup\@href {#1}{\urlprefix }}%
\providecommand \urlprefix  [0]{URL }%
\providecommand \Eprint [0]{\href }%
\providecommand \doibase [0]{https://doi.org/}%
\providecommand \selectlanguage [0]{\@gobble}%
\providecommand \bibinfo  [0]{\@secondoftwo}%
\providecommand \bibfield  [0]{\@secondoftwo}%
\providecommand \translation [1]{[#1]}%
\providecommand \BibitemOpen [0]{}%
\providecommand \bibitemStop [0]{}%
\providecommand \bibitemNoStop [0]{.\EOS\space}%
\providecommand \EOS [0]{\spacefactor3000\relax}%
\providecommand \BibitemShut  [1]{\csname bibitem#1\endcsname}%
\let\auto@bib@innerbib\@empty
\bibitem [{\citenamefont {Adamian}\ \emph {et~al.}(2020)\citenamefont
  {Adamian}, \citenamefont {Antonenko}, \citenamefont {Diaz-Torres},\ and\
  \citenamefont {Heinz}}]{adamian2020}%
  \BibitemOpen
  \bibfield  {author} {\bibinfo {author} {\bibfnamefont {G.~G.}\ \bibnamefont
  {Adamian}}, \bibinfo {author} {\bibfnamefont {N.~V.}\ \bibnamefont
  {Antonenko}}, \bibinfo {author} {\bibfnamefont {A.}~\bibnamefont
  {Diaz-Torres}},\ and\ \bibinfo {author} {\bibfnamefont {S.}~\bibnamefont
  {Heinz}},\ }\bibfield  {title} {\bibinfo {title} {How to extend the chart of
  nuclides?},\ }\href {https://doi.org/10.1140/epja/s10050-020-00046-7}
  {\bibfield  {journal} {\bibinfo  {journal} {Eur. Phys. J. A}\ }\textbf
  {\bibinfo {volume} {56}},\ \bibinfo {pages} {47} (\bibinfo {year}
  {2020})}\BibitemShut {NoStop}%
\bibitem [{\citenamefont {M\"unzenberg}\ \emph {et~al.}(2023)\citenamefont
  {M\"unzenberg}, \citenamefont {Gupta}, \citenamefont {Devaraja},
  \citenamefont {Gambhir}, \citenamefont {Heinz},\ and\ \citenamefont
  {Hofmann}}]{munzenberg2023}%
  \BibitemOpen
  \bibfield  {author} {\bibinfo {author} {\bibfnamefont {G.}~\bibnamefont
  {M\"unzenberg}}, \bibinfo {author} {\bibfnamefont {M.}~\bibnamefont {Gupta}},
  \bibinfo {author} {\bibfnamefont {H.~M.}\ \bibnamefont {Devaraja}}, \bibinfo
  {author} {\bibfnamefont {Y.~K.}\ \bibnamefont {Gambhir}}, \bibinfo {author}
  {\bibfnamefont {S.}~\bibnamefont {Heinz}},\ and\ \bibinfo {author}
  {\bibfnamefont {S.}~\bibnamefont {Hofmann}},\ }\bibfield  {title} {\bibinfo
  {title} {Heavy and superheavy elements: next generation experiments, ideas
  and considerations},\ }\href
  {https://doi.org/10.1140/epja/s10050-023-00939-3} {\bibfield  {journal}
  {\bibinfo  {journal} {Eur. Phys. J. A}\ }\textbf {\bibinfo {volume} {59}},\
  \bibinfo {pages} {21} (\bibinfo {year} {2023})}\BibitemShut {NoStop}%
\bibitem [{\citenamefont {Niwase}\ \emph {et~al.}(2023)\citenamefont {Niwase},
  \citenamefont {Watanabe}, \citenamefont {Hirayama}, \citenamefont {Mukai},
  \citenamefont {Schury}, \citenamefont {Andreyev}, \citenamefont {Hashimoto},
  \citenamefont {Iimura}, \citenamefont {Ishiyama}, \citenamefont {Ito},
  \citenamefont {Jeong}, \citenamefont {Kaji}, \citenamefont {Kimura},
  \citenamefont {Miyatake}, \citenamefont {Morimoto}, \citenamefont {Moon},
  \citenamefont {Oyaizu}, \citenamefont {Rosenbusch}, \citenamefont
  {Taniguchi},\ and\ \citenamefont {Wada}}]{niwase2023}%
  \BibitemOpen
  \bibfield  {author} {\bibinfo {author} {\bibfnamefont {T.}~\bibnamefont
  {Niwase}}, \bibinfo {author} {\bibfnamefont {Y.~X.}\ \bibnamefont
  {Watanabe}}, \bibinfo {author} {\bibfnamefont {Y.}~\bibnamefont {Hirayama}},
  \bibinfo {author} {\bibfnamefont {M.}~\bibnamefont {Mukai}}, \bibinfo
  {author} {\bibfnamefont {P.}~\bibnamefont {Schury}}, \bibinfo {author}
  {\bibfnamefont {A.~N.}\ \bibnamefont {Andreyev}}, \bibinfo {author}
  {\bibfnamefont {T.}~\bibnamefont {Hashimoto}}, \bibinfo {author}
  {\bibfnamefont {S.}~\bibnamefont {Iimura}}, \bibinfo {author} {\bibfnamefont
  {H.}~\bibnamefont {Ishiyama}}, \bibinfo {author} {\bibfnamefont
  {Y.}~\bibnamefont {Ito}}, \bibinfo {author} {\bibfnamefont {S.~C.}\
  \bibnamefont {Jeong}}, \bibinfo {author} {\bibfnamefont {D.}~\bibnamefont
  {Kaji}}, \bibinfo {author} {\bibfnamefont {S.}~\bibnamefont {Kimura}},
  \bibinfo {author} {\bibfnamefont {H.}~\bibnamefont {Miyatake}}, \bibinfo
  {author} {\bibfnamefont {K.}~\bibnamefont {Morimoto}}, \bibinfo {author}
  {\bibfnamefont {J.-Y.}\ \bibnamefont {Moon}}, \bibinfo {author}
  {\bibfnamefont {M.}~\bibnamefont {Oyaizu}}, \bibinfo {author} {\bibfnamefont
  {M.}~\bibnamefont {Rosenbusch}}, \bibinfo {author} {\bibfnamefont
  {A.}~\bibnamefont {Taniguchi}},\ and\ \bibinfo {author} {\bibfnamefont
  {M.}~\bibnamefont {Wada}},\ }\bibfield  {title} {\bibinfo {title} {Discovery
  of {N}ew {I}sotope $^{241}\mathrm{U}$ and {S}ystematic {H}igh-{P}recision
  {A}tomic {M}ass {M}easurements of {N}eutron-{R}ich {Pa-Pu} {N}uclei
  {P}roduced via {M}ultinucleon {T}ransfer {R}eactions},\ }\href
  {https://doi.org/10.1103/PhysRevLett.130.132502} {\bibfield  {journal}
  {\bibinfo  {journal} {Phys. Rev. Lett.}\ }\textbf {\bibinfo {volume} {130}},\
  \bibinfo {pages} {132502} (\bibinfo {year} {2023})}\BibitemShut {NoStop}%
\bibitem [{\citenamefont {{Valery Zagrebaev}}\ and\ \citenamefont {{Walter
  Greiner}}(2008{\natexlab{a}})}]{zagrebaev2008}%
  \BibitemOpen
  \bibfield  {author} {\bibinfo {author} {\bibnamefont {{Valery Zagrebaev}}}\
  and\ \bibinfo {author} {\bibnamefont {{Walter Greiner}}},\ }\bibfield
  {title} {\bibinfo {title} {{N}ew way for the production of heavy neutron-rich
  nuclei},\ }\href {https://doi.org/10.1088/0954-3899/35/12/125103} {\bibfield
  {journal} {\bibinfo  {journal} {J. Phys. G: Nucl. Part. Phys.}\ }\textbf
  {\bibinfo {volume} {35}},\ \bibinfo {pages} {125103} (\bibinfo {year}
  {2008}{\natexlab{a}})}\BibitemShut {NoStop}%
\bibitem [{\citenamefont {{Valery Zagrebaev}}\ and\ \citenamefont {{Walter
  Greiner}}(2008{\natexlab{b}})}]{zagrebaev2008c}%
  \BibitemOpen
  \bibfield  {author} {\bibinfo {author} {\bibnamefont {{Valery Zagrebaev}}}\
  and\ \bibinfo {author} {\bibnamefont {{Walter Greiner}}},\ }\bibfield
  {title} {\bibinfo {title} {Production of {N}ew {H}eavy {I}sotopes in
  {L}ow--{E}nergy {M}ultinucleon {T}ransfer {R}eactions},\ }\href
  {https://doi.org/10.1103/PhysRevLett.101.122701} {\bibfield  {journal}
  {\bibinfo  {journal} {Phys. Rev. Lett.}\ }\textbf {\bibinfo {volume} {101}},\
  \bibinfo {pages} {122701} (\bibinfo {year} {2008}{\natexlab{b}})}\BibitemShut
  {NoStop}%
\bibitem [{\citenamefont {Zagrebaev}\ and\ \citenamefont
  {Greiner}(2011)}]{zagrebaev2011}%
  \BibitemOpen
  \bibfield  {author} {\bibinfo {author} {\bibfnamefont {V.~I.}\ \bibnamefont
  {Zagrebaev}}\ and\ \bibinfo {author} {\bibfnamefont {W.}~\bibnamefont
  {Greiner}},\ }\bibfield  {title} {\bibinfo {title} {{P}roduction of heavy and
  superheavy neutron-rich nuclei in transfer reactions},\ }\href
  {https://doi.org/10.1103/PhysRevC.83.044618} {\bibfield  {journal} {\bibinfo
  {journal} {Phys. Rev. C}\ }\textbf {\bibinfo {volume} {83}},\ \bibinfo
  {pages} {044618} (\bibinfo {year} {2011})}\BibitemShut {NoStop}%
\bibitem [{\citenamefont {Karpov}\ and\ \citenamefont
  {Saiko}(2017)}]{karpov2017}%
  \BibitemOpen
  \bibfield  {author} {\bibinfo {author} {\bibfnamefont {A.~V.}\ \bibnamefont
  {Karpov}}\ and\ \bibinfo {author} {\bibfnamefont {V.~V.}\ \bibnamefont
  {Saiko}},\ }\bibfield  {title} {\bibinfo {title} {Modeling near-barrier
  collisions of heavy ions based on a {L}angevin-type approach},\ }\href
  {https://doi.org/10.1103/PhysRevC.96.024618} {\bibfield  {journal} {\bibinfo
  {journal} {Phys. Rev. C}\ }\textbf {\bibinfo {volume} {96}},\ \bibinfo
  {pages} {024618} (\bibinfo {year} {2017})}\BibitemShut {NoStop}%
\bibitem [{\citenamefont {Saiko}\ and\ \citenamefont
  {Karpov}(2019)}]{saiko2019}%
  \BibitemOpen
  \bibfield  {author} {\bibinfo {author} {\bibfnamefont {V.~V.}\ \bibnamefont
  {Saiko}}\ and\ \bibinfo {author} {\bibfnamefont {A.~V.}\ \bibnamefont
  {Karpov}},\ }\bibfield  {title} {\bibinfo {title} {Analysis of multinucleon
  transfer reactions with spherical and statically deformed nuclei using a
  {L}angevin-type approach},\ }\href
  {https://doi.org/10.1103/PhysRevC.99.014613} {\bibfield  {journal} {\bibinfo
  {journal} {Phys. Rev. C}\ }\textbf {\bibinfo {volume} {99}},\ \bibinfo
  {pages} {014613} (\bibinfo {year} {2019})}\BibitemShut {NoStop}%
\bibitem [{\citenamefont {Saiko}\ and\ \citenamefont
  {Karpov}(2022)}]{saiko2022}%
  \BibitemOpen
  \bibfield  {author} {\bibinfo {author} {\bibfnamefont {V.}~\bibnamefont
  {Saiko}}\ and\ \bibinfo {author} {\bibfnamefont {A.}~\bibnamefont {Karpov}},\
  }\bibfield  {title} {\bibinfo {title} {Multinucleon transfer as a method for
  production of new heavy neutron-enriched isotopes of transuranium elements},\
  }\href {https://doi.org/10.1140/epja/s10050-022-00688-9} {\bibfield
  {journal} {\bibinfo  {journal} {Eur. Phys. J. A}\ }\textbf {\bibinfo {volume}
  {58}},\ \bibinfo {pages} {41} (\bibinfo {year} {2022})}\BibitemShut {NoStop}%
\bibitem [{\citenamefont {Feng}\ \emph {et~al.}(2009)\citenamefont {Feng},
  \citenamefont {Jin},\ and\ \citenamefont {Li}}]{feng2009a}%
  \BibitemOpen
  \bibfield  {author} {\bibinfo {author} {\bibfnamefont {Z.-Q.}\ \bibnamefont
  {Feng}}, \bibinfo {author} {\bibfnamefont {G.-M.}\ \bibnamefont {Jin}},\ and\
  \bibinfo {author} {\bibfnamefont {J.-Q.}\ \bibnamefont {Li}},\ }\bibfield
  {title} {\bibinfo {title} {Production of heavy isotopes in transfer reactions
  by collisions of $^{238}\mathrm{U}+{}^{238}\mathrm{U}$},\ }\href
  {https://doi.org/10.1103/PhysRevC.80.067601} {\bibfield  {journal} {\bibinfo
  {journal} {Phys. Rev. C}\ }\textbf {\bibinfo {volume} {80}},\ \bibinfo
  {pages} {067601} (\bibinfo {year} {2009})}\BibitemShut {NoStop}%
\bibitem [{\citenamefont {Feng}(2017)}]{feng2017}%
  \BibitemOpen
  \bibfield  {author} {\bibinfo {author} {\bibfnamefont {Z.-Q.}\ \bibnamefont
  {Feng}},\ }\bibfield  {title} {\bibinfo {title} {Production of neutron--rich
  isotopes around ${N}=126$ in multinucleon transfer reactions},\ }\href
  {https://doi.org/10.1103/PhysRevC.95.024615} {\bibfield  {journal} {\bibinfo
  {journal} {Phys. Rev. C}\ }\textbf {\bibinfo {volume} {95}},\ \bibinfo
  {pages} {024615} (\bibinfo {year} {2017})}\BibitemShut {NoStop}%
\bibitem [{\citenamefont {Zhao}\ \emph {et~al.}(2009)\citenamefont {Zhao},
  \citenamefont {Wu},\ and\ \citenamefont {Li}}]{zhao2009}%
  \BibitemOpen
  \bibfield  {author} {\bibinfo {author} {\bibfnamefont {K.}~\bibnamefont
  {Zhao}}, \bibinfo {author} {\bibfnamefont {X.}~\bibnamefont {Wu}},\ and\
  \bibinfo {author} {\bibfnamefont {Z.}~\bibnamefont {Li}},\ }\bibfield
  {title} {\bibinfo {title} {Quantum molecular dynamics study of the mass
  distribution of products in $7.0{A}$\,{M}e{V}
  $^{238}\mathrm{U}+{}^{238}\mathrm{U}$ collisions},\ }\href
  {https://doi.org/10.1103/PhysRevC.80.054607} {\bibfield  {journal} {\bibinfo
  {journal} {Phys. Rev. C}\ }\textbf {\bibinfo {volume} {80}},\ \bibinfo
  {pages} {054607} (\bibinfo {year} {2009})}\BibitemShut {NoStop}%
\bibitem [{\citenamefont {Zhao}\ \emph {et~al.}(2016)\citenamefont {Zhao},
  \citenamefont {Li}, \citenamefont {Zhang}, \citenamefont {Wang},
  \citenamefont {Li}, \citenamefont {Shen}, \citenamefont {Wang},\ and\
  \citenamefont {Wu}}]{zhao2016}%
  \BibitemOpen
  \bibfield  {author} {\bibinfo {author} {\bibfnamefont {K.}~\bibnamefont
  {Zhao}}, \bibinfo {author} {\bibfnamefont {Z.}~\bibnamefont {Li}}, \bibinfo
  {author} {\bibfnamefont {Y.}~\bibnamefont {Zhang}}, \bibinfo {author}
  {\bibfnamefont {N.}~\bibnamefont {Wang}}, \bibinfo {author} {\bibfnamefont
  {Q.}~\bibnamefont {Li}}, \bibinfo {author} {\bibfnamefont {C.}~\bibnamefont
  {Shen}}, \bibinfo {author} {\bibfnamefont {Y.}~\bibnamefont {Wang}},\ and\
  \bibinfo {author} {\bibfnamefont {X.}~\bibnamefont {Wu}},\ }\bibfield
  {title} {\bibinfo {title} {Production of unknown neutron--rich isotopes in
  $^{238}\mathrm{U}+{}^{238}\mathrm{U}$ collisions at near--barrier energy},\
  }\href {https://doi.org/10.1103/PhysRevC.94.024601} {\bibfield  {journal}
  {\bibinfo  {journal} {Phys. Rev. C}\ }\textbf {\bibinfo {volume} {94}},\
  \bibinfo {pages} {024601} (\bibinfo {year} {2016})}\BibitemShut {NoStop}%
\bibitem [{\citenamefont {Wang}\ and\ \citenamefont {Guo}(2016)}]{wang2016}%
  \BibitemOpen
  \bibfield  {author} {\bibinfo {author} {\bibfnamefont {N.}~\bibnamefont
  {Wang}}\ and\ \bibinfo {author} {\bibfnamefont {L.}~\bibnamefont {Guo}},\
  }\bibfield  {title} {\bibinfo {title} {New neutron-rich isotope production in
  $^{154}\mathrm{Sm}+{}^{160}\mathrm{Gd}$},\ }\href
  {https://doi.org/10.1016/j.physletb.2016.06.073} {\bibfield  {journal}
  {\bibinfo  {journal} {Phys. Lett. B}\ }\textbf {\bibinfo {volume} {760}},\
  \bibinfo {pages} {236} (\bibinfo {year} {2016})}\BibitemShut {NoStop}%
\bibitem [{\citenamefont {Simenel}(2012)}]{simenel2012}%
  \BibitemOpen
  \bibfield  {author} {\bibinfo {author} {\bibfnamefont {C.}~\bibnamefont
  {Simenel}},\ }\bibfield  {title} {\bibinfo {title} {{N}uclear quantum
  many-body dynamics},\ }\href {https://doi.org/10.1140/epja/i2012-12152-0}
  {\bibfield  {journal} {\bibinfo  {journal} {Eur. Phys. J. A}\ }\textbf
  {\bibinfo {volume} {48}},\ \bibinfo {pages} {152} (\bibinfo {year}
  {2012})}\BibitemShut {NoStop}%
\bibitem [{\citenamefont {Simenel}\ and\ \citenamefont
  {Umar}(2018)}]{simenel2018}%
  \BibitemOpen
  \bibfield  {author} {\bibinfo {author} {\bibfnamefont {C.}~\bibnamefont
  {Simenel}}\ and\ \bibinfo {author} {\bibfnamefont {A.~S.}\ \bibnamefont
  {Umar}},\ }\bibfield  {title} {\bibinfo {title} {Heavy-ion collisions and
  fission dynamics with the time--dependent {H}artree-{F}ock theory and its
  extensions},\ }\href {https://doi.org/10.1016/j.ppnp.2018.07.002} {\bibfield
  {journal} {\bibinfo  {journal} {Prog. Part. Nucl. Phys.}\ }\textbf {\bibinfo
  {volume} {103}},\ \bibinfo {pages} {19} (\bibinfo {year} {2018})}\BibitemShut
  {NoStop}%
\bibitem [{\citenamefont {Nakatsukasa}\ \emph {et~al.}(2016)\citenamefont
  {Nakatsukasa}, \citenamefont {Matsuyanagi}, \citenamefont {Matsuo},\ and\
  \citenamefont {Yabana}}]{nakatsukasa2016}%
  \BibitemOpen
  \bibfield  {author} {\bibinfo {author} {\bibfnamefont {T.}~\bibnamefont
  {Nakatsukasa}}, \bibinfo {author} {\bibfnamefont {K.}~\bibnamefont
  {Matsuyanagi}}, \bibinfo {author} {\bibfnamefont {M.}~\bibnamefont
  {Matsuo}},\ and\ \bibinfo {author} {\bibfnamefont {K.}~\bibnamefont
  {Yabana}},\ }\bibfield  {title} {\bibinfo {title} {Time-dependent
  density-functional description of nuclear dynamics},\ }\href
  {https://doi.org/10.1103/RevModPhys.88.045004} {\bibfield  {journal}
  {\bibinfo  {journal} {Rev. Mod. Phys.}\ }\textbf {\bibinfo {volume} {88}},\
  \bibinfo {pages} {045004} (\bibinfo {year} {2016})}\BibitemShut {NoStop}%
\bibitem [{\citenamefont {Oberacker}\ \emph {et~al.}(2014)\citenamefont
  {Oberacker}, \citenamefont {Umar},\ and\ \citenamefont
  {Simenel}}]{oberacker2014}%
  \BibitemOpen
  \bibfield  {author} {\bibinfo {author} {\bibfnamefont {V.~E.}\ \bibnamefont
  {Oberacker}}, \bibinfo {author} {\bibfnamefont {A.~S.}\ \bibnamefont
  {Umar}},\ and\ \bibinfo {author} {\bibfnamefont {C.}~\bibnamefont
  {Simenel}},\ }\bibfield  {title} {\bibinfo {title} {{D}issipative dynamics in
  quasifission},\ }\href {https://doi.org/10.1103/PhysRevC.90.054605}
  {\bibfield  {journal} {\bibinfo  {journal} {Phys. Rev. C}\ }\textbf {\bibinfo
  {volume} {90}},\ \bibinfo {pages} {054605} (\bibinfo {year}
  {2014})}\BibitemShut {NoStop}%
\bibitem [{\citenamefont {Umar}\ \emph {et~al.}(2015)\citenamefont {Umar},
  \citenamefont {Oberacker},\ and\ \citenamefont {Simenel}}]{umar2015a}%
  \BibitemOpen
  \bibfield  {author} {\bibinfo {author} {\bibfnamefont {A.~S.}\ \bibnamefont
  {Umar}}, \bibinfo {author} {\bibfnamefont {V.~E.}\ \bibnamefont
  {Oberacker}},\ and\ \bibinfo {author} {\bibfnamefont {C.}~\bibnamefont
  {Simenel}},\ }\bibfield  {title} {\bibinfo {title} {{S}hape evolution and
  collective dynamics of quasifission in the time-dependent {H}artree-{F}ock
  approach},\ }\href {https://doi.org/10.1103/PhysRevC.92.024621} {\bibfield
  {journal} {\bibinfo  {journal} {Phys. Rev. C}\ }\textbf {\bibinfo {volume}
  {92}},\ \bibinfo {pages} {024621} (\bibinfo {year} {2015})}\BibitemShut
  {NoStop}%
\bibitem [{\citenamefont {Umar}\ and\ \citenamefont
  {Oberacker}(2015)}]{umar2015c}%
  \BibitemOpen
  \bibfield  {author} {\bibinfo {author} {\bibfnamefont {A.~S.}\ \bibnamefont
  {Umar}}\ and\ \bibinfo {author} {\bibfnamefont {V.~E.}\ \bibnamefont
  {Oberacker}},\ }\bibfield  {title} {\bibinfo {title} {{T}ime-dependent {HF}
  approach to {SHE} dynamics},\ }\href
  {https://doi.org/10.1016/j.nuclphysa.2015.02.011} {\bibfield  {journal}
  {\bibinfo  {journal} {Nucl. Phys. A}\ }\textbf {\bibinfo {volume} {944}},\
  \bibinfo {pages} {238} (\bibinfo {year} {2015})}\BibitemShut {NoStop}%
\bibitem [{\citenamefont {Umar}\ \emph {et~al.}(2017)\citenamefont {Umar},
  \citenamefont {Simenel},\ and\ \citenamefont {Ye}}]{umar2017}%
  \BibitemOpen
  \bibfield  {author} {\bibinfo {author} {\bibfnamefont {A.~S.}\ \bibnamefont
  {Umar}}, \bibinfo {author} {\bibfnamefont {C.}~\bibnamefont {Simenel}},\ and\
  \bibinfo {author} {\bibfnamefont {W.}~\bibnamefont {Ye}},\ }\bibfield
  {title} {\bibinfo {title} {Transport properties of isospin asymmetric nuclear
  matter using the time-dependent {H}artree--{F}ock method},\ }\href
  {https://doi.org/10.1103/PhysRevC.96.024625} {\bibfield  {journal} {\bibinfo
  {journal} {Phys. Rev. C}\ }\textbf {\bibinfo {volume} {96}},\ \bibinfo
  {pages} {024625} (\bibinfo {year} {2017})}\BibitemShut {NoStop}%
\bibitem [{\citenamefont {Simenel}(2010)}]{simenel2010}%
  \BibitemOpen
  \bibfield  {author} {\bibinfo {author} {\bibfnamefont {C.}~\bibnamefont
  {Simenel}},\ }\bibfield  {title} {\bibinfo {title} {{P}article {T}ransfer
  {R}eactions with the {T}ime-{D}ependent {H}artree-{F}ock {T}heory {U}sing a
  {P}article {N}umber {P}rojection {T}echnique},\ }\href
  {https://doi.org/10.1103/PhysRevLett.105.192701} {\bibfield  {journal}
  {\bibinfo  {journal} {Phys. Rev. Lett.}\ }\textbf {\bibinfo {volume} {105}},\
  \bibinfo {pages} {192701} (\bibinfo {year} {2010})}\BibitemShut {NoStop}%
\bibitem [{\citenamefont {Sekizawa}\ and\ \citenamefont
  {Yabana}(2016)}]{sekizawa2016}%
  \BibitemOpen
  \bibfield  {author} {\bibinfo {author} {\bibfnamefont {K.}~\bibnamefont
  {Sekizawa}}\ and\ \bibinfo {author} {\bibfnamefont {K.}~\bibnamefont
  {Yabana}},\ }\bibfield  {title} {\bibinfo {title} {{T}ime-dependent
  {H}artree-{F}ock calculations for multinucleon transfer and quasifission
  processes in the $^{64}\mathrm{Ni}+{}^{238}\mathrm{U}$ reaction},\ }\href
  {https://doi.org/10.1103/PhysRevC.93.054616} {\bibfield  {journal} {\bibinfo
  {journal} {Phys. Rev. C}\ }\textbf {\bibinfo {volume} {93}},\ \bibinfo
  {pages} {054616} (\bibinfo {year} {2016})}\BibitemShut {NoStop}%
\bibitem [{\citenamefont {Godbey}\ \emph {et~al.}(2019)\citenamefont {Godbey},
  \citenamefont {Umar},\ and\ \citenamefont {Simenel}}]{godbey2019}%
  \BibitemOpen
  \bibfield  {author} {\bibinfo {author} {\bibfnamefont {K.}~\bibnamefont
  {Godbey}}, \bibinfo {author} {\bibfnamefont {A.~S.}\ \bibnamefont {Umar}},\
  and\ \bibinfo {author} {\bibfnamefont {C.}~\bibnamefont {Simenel}},\
  }\bibfield  {title} {\bibinfo {title} {Deformed shell effects in
  ${}^{48}\mathrm{Ca}+{}^{249}\mathrm{Bk}$ quasifission fragments},\ }\href
  {https://doi.org/10.1103/PhysRevC.100.024610} {\bibfield  {journal} {\bibinfo
   {journal} {Phys. Rev. C}\ }\textbf {\bibinfo {volume} {100}},\ \bibinfo
  {pages} {024610} (\bibinfo {year} {2019})}\BibitemShut {NoStop}%
\bibitem [{\citenamefont {Simenel}\ \emph {et~al.}(2020)\citenamefont
  {Simenel}, \citenamefont {Godbey},\ and\ \citenamefont {Umar}}]{simenel2020}%
  \BibitemOpen
  \bibfield  {author} {\bibinfo {author} {\bibfnamefont {C.}~\bibnamefont
  {Simenel}}, \bibinfo {author} {\bibfnamefont {K.}~\bibnamefont {Godbey}},\
  and\ \bibinfo {author} {\bibfnamefont {A.~S.}\ \bibnamefont {Umar}},\
  }\bibfield  {title} {\bibinfo {title} {Timescales of {Q}uantum
  {E}quilibration, {D}issipation and {F}luctuation in {N}uclear {C}ollisions},\
  }\href {https://doi.org/10.1103/PhysRevLett.124.212504} {\bibfield  {journal}
  {\bibinfo  {journal} {Phys. Rev. Lett.}\ }\textbf {\bibinfo {volume} {124}},\
  \bibinfo {pages} {212504} (\bibinfo {year} {2020})}\BibitemShut {NoStop}%
\bibitem [{\citenamefont {Balian}\ and\ \citenamefont
  {V\'en\'eroni}(1985)}]{balian1985}%
  \BibitemOpen
  \bibfield  {author} {\bibinfo {author} {\bibfnamefont {R.}~\bibnamefont
  {Balian}}\ and\ \bibinfo {author} {\bibfnamefont {M.}~\bibnamefont
  {V\'en\'eroni}},\ }\bibfield  {title} {\bibinfo {title} {Time-dependent
  variational principle for the expectation value of an observable:
  {M}ean-field applications},\ }\href
  {https://doi.org/10.1016/0003-4916(85)90020-X} {\bibfield  {journal}
  {\bibinfo  {journal} {Ann. Phys. {(NY)}}\ }\textbf {\bibinfo {volume}
  {164}},\ \bibinfo {pages} {334} (\bibinfo {year} {1985})}\BibitemShut
  {NoStop}%
\bibitem [{\citenamefont {Balian}\ and\ \citenamefont
  {V\'en\'eroni}(1992)}]{balian1992}%
  \BibitemOpen
  \bibfield  {author} {\bibinfo {author} {\bibfnamefont {R.}~\bibnamefont
  {Balian}}\ and\ \bibinfo {author} {\bibfnamefont {M.}~\bibnamefont
  {V\'en\'eroni}},\ }\bibfield  {title} {\bibinfo {title} {Correlations and
  fluctuations in static and dynamic mean-field approaches},\ }\href
  {https://doi.org/10.1016/0003-4916(92)90181-K} {\bibfield  {journal}
  {\bibinfo  {journal} {Ann. Phys.}\ }\textbf {\bibinfo {volume} {216}},\
  \bibinfo {pages} {351} (\bibinfo {year} {1992})}\BibitemShut {NoStop}%
\bibitem [{\citenamefont {Simenel}(2011)}]{simenel2011}%
  \BibitemOpen
  \bibfield  {author} {\bibinfo {author} {\bibfnamefont {C.}~\bibnamefont
  {Simenel}},\ }\bibfield  {title} {\bibinfo {title} {{P}article-{N}umber
  {F}luctuations and {C}orrelations in {T}ransfer {R}eactions {O}btained
  {U}sing the {B}alian-{V}\'en\'eroni {V}ariational {P}rinciple},\ }\href
  {https://doi.org/10.1103/PhysRevLett.106.112502} {\bibfield  {journal}
  {\bibinfo  {journal} {Phys. Rev. Lett.}\ }\textbf {\bibinfo {volume} {106}},\
  \bibinfo {pages} {112502} (\bibinfo {year} {2011})}\BibitemShut {NoStop}%
\bibitem [{\citenamefont {Williams}\ \emph {et~al.}(2018)\citenamefont
  {Williams}, \citenamefont {Sekizawa}, \citenamefont {Hinde}, \citenamefont
  {Simenel}, \citenamefont {Dasgupta}, \citenamefont {Carter}, \citenamefont
  {Cook}, \citenamefont {Jeung}, \citenamefont {McNeil}, \citenamefont
  {Palshetkar}, \citenamefont {Rafferty}, \citenamefont {Ramachandran},\ and\
  \citenamefont {Wakhle}}]{williams2018}%
  \BibitemOpen
  \bibfield  {author} {\bibinfo {author} {\bibfnamefont {E.}~\bibnamefont
  {Williams}}, \bibinfo {author} {\bibfnamefont {K.}~\bibnamefont {Sekizawa}},
  \bibinfo {author} {\bibfnamefont {D.~J.}\ \bibnamefont {Hinde}}, \bibinfo
  {author} {\bibfnamefont {C.}~\bibnamefont {Simenel}}, \bibinfo {author}
  {\bibfnamefont {M.}~\bibnamefont {Dasgupta}}, \bibinfo {author}
  {\bibfnamefont {I.~P.}\ \bibnamefont {Carter}}, \bibinfo {author}
  {\bibfnamefont {K.~J.}\ \bibnamefont {Cook}}, \bibinfo {author}
  {\bibfnamefont {D.~Y.}\ \bibnamefont {Jeung}}, \bibinfo {author}
  {\bibfnamefont {S.~D.}\ \bibnamefont {McNeil}}, \bibinfo {author}
  {\bibfnamefont {C.~S.}\ \bibnamefont {Palshetkar}}, \bibinfo {author}
  {\bibfnamefont {D.~C.}\ \bibnamefont {Rafferty}}, \bibinfo {author}
  {\bibfnamefont {K.}~\bibnamefont {Ramachandran}},\ and\ \bibinfo {author}
  {\bibfnamefont {A.}~\bibnamefont {Wakhle}},\ }\bibfield  {title} {\bibinfo
  {title} {Exploring {Z}eptosecond {Q}uantum {E}quilibration {D}ynamics: {F}rom
  {D}eep-{I}nelastic to {F}usion-{F}ission {O}utcomes in
  $^{58}\mathrm{Ni}+{}^{60}\mathrm{Ni}$ {R}eactions},\ }\href
  {https://doi.org/10.1103/PhysRevLett.120.022501} {\bibfield  {journal}
  {\bibinfo  {journal} {Phys. Rev. Lett.}\ }\textbf {\bibinfo {volume} {120}},\
  \bibinfo {pages} {022501} (\bibinfo {year} {2018})}\BibitemShut {NoStop}%
\bibitem [{\citenamefont {Godbey}\ and\ \citenamefont
  {Umar}(2020)}]{godbey2020}%
  \BibitemOpen
  \bibfield  {author} {\bibinfo {author} {\bibfnamefont {K.}~\bibnamefont
  {Godbey}}\ and\ \bibinfo {author} {\bibfnamefont {A.~S.}\ \bibnamefont
  {Umar}},\ }\bibfield  {title} {\bibinfo {title} {Quasifission {D}ynamics in
  {M}icroscopic {T}heories},\ }\href {https://doi.org/10.3389/fphy.2020.00040}
  {\bibfield  {journal} {\bibinfo  {journal} {Front. Phys.}\ }\textbf {\bibinfo
  {volume} {8}},\ \bibinfo {pages} {40} (\bibinfo {year} {2020})}\BibitemShut
  {NoStop}%
\bibitem [{\citenamefont {Godbey}\ \emph {et~al.}(2020)\citenamefont {Godbey},
  \citenamefont {Simenel},\ and\ \citenamefont {Umar}}]{godbey2020b}%
  \BibitemOpen
  \bibfield  {author} {\bibinfo {author} {\bibfnamefont {K.}~\bibnamefont
  {Godbey}}, \bibinfo {author} {\bibfnamefont {C.}~\bibnamefont {Simenel}},\
  and\ \bibinfo {author} {\bibfnamefont {A.~S.}\ \bibnamefont {Umar}},\
  }\bibfield  {title} {\bibinfo {title} {Microscopic predictions for the
  production of neutron-rich nuclei in the reaction
  $^{176}\mathrm{Yb}+{}^{176}\mathrm{Yb}$},\ }\href
  {https://doi.org/10.1103/PhysRevC.101.034602} {\bibfield  {journal} {\bibinfo
   {journal} {Phys. Rev. C}\ }\textbf {\bibinfo {volume} {101}},\ \bibinfo
  {pages} {034602} (\bibinfo {year} {2020})}\BibitemShut {NoStop}%
\bibitem [{\citenamefont {Ayik}(2008)}]{ayik2008}%
  \BibitemOpen
  \bibfield  {author} {\bibinfo {author} {\bibfnamefont {S.}~\bibnamefont
  {Ayik}},\ }\bibfield  {title} {\bibinfo {title} {A stochastic mean-field
  approach for nuclear dynamics},\ }\href
  {https://doi.org/10.1016/j.physletb.2007.09.072} {\bibfield  {journal}
  {\bibinfo  {journal} {Phys. Lett. B}\ }\textbf {\bibinfo {volume} {658}},\
  \bibinfo {pages} {174} (\bibinfo {year} {2008})}\BibitemShut {NoStop}%
\bibitem [{\citenamefont {Lacroix}\ and\ \citenamefont
  {Ayik}(2014)}]{lacroix2014}%
  \BibitemOpen
  \bibfield  {author} {\bibinfo {author} {\bibfnamefont {D.}~\bibnamefont
  {Lacroix}}\ and\ \bibinfo {author} {\bibfnamefont {S.}~\bibnamefont {Ayik}},\
  }\bibfield  {title} {\bibinfo {title} {{S}tochastic quantum dynamics beyond
  mean field},\ }\href {https://doi.org/10.1140/epja/i2014-14095-8} {\bibfield
  {journal} {\bibinfo  {journal} {Eur. Phys. J. A}\ }\textbf {\bibinfo {volume}
  {50}},\ \bibinfo {pages} {95} (\bibinfo {year} {2014})}\BibitemShut {NoStop}%
\bibitem [{\citenamefont {Ayik}\ \emph {et~al.}(2018)\citenamefont {Ayik},
  \citenamefont {Yilmaz}, \citenamefont {Yilmaz},\ and\ \citenamefont
  {Umar}}]{ayik2018}%
  \BibitemOpen
  \bibfield  {author} {\bibinfo {author} {\bibfnamefont {S.}~\bibnamefont
  {Ayik}}, \bibinfo {author} {\bibfnamefont {B.}~\bibnamefont {Yilmaz}},
  \bibinfo {author} {\bibfnamefont {O.}~\bibnamefont {Yilmaz}},\ and\ \bibinfo
  {author} {\bibfnamefont {A.~S.}\ \bibnamefont {Umar}},\ }\bibfield  {title}
  {\bibinfo {title} {Quantal diffusion description of multinucleon transfers in
  heavy--ion collisions},\ }\href {https://doi.org/10.1103/PhysRevC.97.054618}
  {\bibfield  {journal} {\bibinfo  {journal} {Phys. Rev. C}\ }\textbf {\bibinfo
  {volume} {97}},\ \bibinfo {pages} {054618} (\bibinfo {year}
  {2018})}\BibitemShut {NoStop}%
\bibitem [{\citenamefont {Ayik}\ \emph {et~al.}(2019)\citenamefont {Ayik},
  \citenamefont {Yilmaz}, \citenamefont {Yilmaz},\ and\ \citenamefont
  {Umar}}]{ayik2019}%
  \BibitemOpen
  \bibfield  {author} {\bibinfo {author} {\bibfnamefont {S.}~\bibnamefont
  {Ayik}}, \bibinfo {author} {\bibfnamefont {B.}~\bibnamefont {Yilmaz}},
  \bibinfo {author} {\bibfnamefont {O.}~\bibnamefont {Yilmaz}},\ and\ \bibinfo
  {author} {\bibfnamefont {A.~S.}\ \bibnamefont {Umar}},\ }\bibfield  {title}
  {\bibinfo {title} {Quantal diffusion approach for multinucleon transfers in
  $\mathrm{Xe}+\mathrm{Pb}$ collisions},\ }\href
  {https://doi.org/10.1103/PhysRevC.100.014609} {\bibfield  {journal} {\bibinfo
   {journal} {Phys. Rev. C}\ }\textbf {\bibinfo {volume} {100}},\ \bibinfo
  {pages} {014609} (\bibinfo {year} {2019})}\BibitemShut {NoStop}%
\bibitem [{\citenamefont {Sekizawa}\ and\ \citenamefont
  {Ayik}(2020)}]{sekizawa2020}%
  \BibitemOpen
  \bibfield  {author} {\bibinfo {author} {\bibfnamefont {K.}~\bibnamefont
  {Sekizawa}}\ and\ \bibinfo {author} {\bibfnamefont {S.}~\bibnamefont
  {Ayik}},\ }\bibfield  {title} {\bibinfo {title} {Quantal diffusion approach
  for multinucleon transfer processes in the
  $^{58,64}\mathrm{Ni}+^{208}\mathrm{Pb}$ reactions: {T}oward the production of
  unknown neutron-rich nuclei},\ }\href
  {https://doi.org/10.1103/PhysRevC.102.014620} {\bibfield  {journal} {\bibinfo
   {journal} {Phys. Rev. C}\ }\textbf {\bibinfo {volume} {102}},\ \bibinfo
  {pages} {014620} (\bibinfo {year} {2020})}\BibitemShut {NoStop}%
\bibitem [{\citenamefont {Yilmaz}\ \emph {et~al.}(2020)\citenamefont {Yilmaz},
  \citenamefont {Turan},\ and\ \citenamefont {Yilmaz}}]{yilmaz2020}%
  \BibitemOpen
  \bibfield  {author} {\bibinfo {author} {\bibfnamefont {O.}~\bibnamefont
  {Yilmaz}}, \bibinfo {author} {\bibfnamefont {G.}~\bibnamefont {Turan}},\ and\
  \bibinfo {author} {\bibfnamefont {B.}~\bibnamefont {Yilmaz}},\ }\bibfield
  {title} {\bibinfo {title} {Quasi-fission and fusion-fission reactions in
  $^{48}\mathrm{Ca}+{}^{208}\mathrm{Pb}$ collisions at ${E}_\mathrm{c.m.}=190$
  {MeV}},\ }\href {https://doi.org/10.1140/epja/s10050-020-00053-8} {\bibfield
  {journal} {\bibinfo  {journal} {Eur. Phys. J. A}\ }\textbf {\bibinfo {volume}
  {56}},\ \bibinfo {pages} {37} (\bibinfo {year} {2020})}\BibitemShut {NoStop}%
\bibitem [{\citenamefont {Ayik}\ \emph {et~al.}(2020)\citenamefont {Ayik},
  \citenamefont {Yilmaz}, \citenamefont {Yilmaz},\ and\ \citenamefont
  {Umar}}]{ayik2020}%
  \BibitemOpen
  \bibfield  {author} {\bibinfo {author} {\bibfnamefont {S.}~\bibnamefont
  {Ayik}}, \bibinfo {author} {\bibfnamefont {B.}~\bibnamefont {Yilmaz}},
  \bibinfo {author} {\bibfnamefont {O.}~\bibnamefont {Yilmaz}},\ and\ \bibinfo
  {author} {\bibfnamefont {A.~S.}\ \bibnamefont {Umar}},\ }\bibfield  {title}
  {\bibinfo {title} {Merging of transport theory with the time-dependent
  {H}artree-{F}ock approach: {M}ultinucleon transfer in $\mathrm{U}+\mathrm{U}$
  collisions},\ }\href {https://doi.org/10.1103/PhysRevC.102.024619} {\bibfield
   {journal} {\bibinfo  {journal} {Phys. Rev. C}\ }\textbf {\bibinfo {volume}
  {102}},\ \bibinfo {pages} {024619} (\bibinfo {year} {2020})}\BibitemShut
  {NoStop}%
\bibitem [{\citenamefont {Ayik}\ \emph {et~al.}(2021)\citenamefont {Ayik},
  \citenamefont {Arik}, \citenamefont {Karanfil}, \citenamefont {Yilmaz},
  \citenamefont {Yilmaz},\ and\ \citenamefont {Umar}}]{ayik2021}%
  \BibitemOpen
  \bibfield  {author} {\bibinfo {author} {\bibfnamefont {S.}~\bibnamefont
  {Ayik}}, \bibinfo {author} {\bibfnamefont {M.}~\bibnamefont {Arik}}, \bibinfo
  {author} {\bibfnamefont {E.~C.}\ \bibnamefont {Karanfil}}, \bibinfo {author}
  {\bibfnamefont {O.}~\bibnamefont {Yilmaz}}, \bibinfo {author} {\bibfnamefont
  {B.}~\bibnamefont {Yilmaz}},\ and\ \bibinfo {author} {\bibfnamefont {A.~S.}\
  \bibnamefont {Umar}},\ }\bibfield  {title} {\bibinfo {title} {Quantal
  diffusion description of isotope production via the multinucleon transfer
  mechanism in $^{48}\mathrm{Ca}+^{238}\mathrm{U}$ collisions},\ }\href
  {https://doi.org/10.1103/PhysRevC.104.054614} {\bibfield  {journal} {\bibinfo
   {journal} {Phys. Rev. C}\ }\textbf {\bibinfo {volume} {104}},\ \bibinfo
  {pages} {054614} (\bibinfo {year} {2021})}\BibitemShut {NoStop}%
\bibitem [{\citenamefont {Ayik}\ \emph
  {et~al.}(2023{\natexlab{a}})\citenamefont {Ayik}, \citenamefont {Arik},
  \citenamefont {Yilmaz}, \citenamefont {Yilmaz},\ and\ \citenamefont
  {Umar}}]{ayik2023}%
  \BibitemOpen
  \bibfield  {author} {\bibinfo {author} {\bibfnamefont {S.}~\bibnamefont
  {Ayik}}, \bibinfo {author} {\bibfnamefont {M.}~\bibnamefont {Arik}}, \bibinfo
  {author} {\bibfnamefont {O.}~\bibnamefont {Yilmaz}}, \bibinfo {author}
  {\bibfnamefont {B.}~\bibnamefont {Yilmaz}},\ and\ \bibinfo {author}
  {\bibfnamefont {A.~S.}\ \bibnamefont {Umar}},\ }\bibfield  {title} {\bibinfo
  {title} {Multinucleon transfer mechanism in
  $^{250}\mathrm{Cf}+^{232}\mathrm{Th}$ collisions using the quantal transport
  description based on the stochastic mean-field approach},\ }\href
  {https://doi.org/10.1103/PhysRevC.107.014609} {\bibfield  {journal} {\bibinfo
   {journal} {Phys. Rev. C}\ }\textbf {\bibinfo {volume} {107}},\ \bibinfo
  {pages} {014609} (\bibinfo {year} {2023}{\natexlab{a}})}\BibitemShut
  {NoStop}%
\bibitem [{\citenamefont {Ayik}\ \emph
  {et~al.}(2023{\natexlab{b}})\citenamefont {Ayik}, \citenamefont {Arik},
  \citenamefont {Erbayri}, \citenamefont {Yilmaz},\ and\ \citenamefont
  {Umar}}]{ayik2023b}%
  \BibitemOpen
  \bibfield  {author} {\bibinfo {author} {\bibfnamefont {S.}~\bibnamefont
  {Ayik}}, \bibinfo {author} {\bibfnamefont {M.}~\bibnamefont {Arik}}, \bibinfo
  {author} {\bibfnamefont {E.}~\bibnamefont {Erbayri}}, \bibinfo {author}
  {\bibfnamefont {O.}~\bibnamefont {Yilmaz}},\ and\ \bibinfo {author}
  {\bibfnamefont {A.~S.}\ \bibnamefont {Umar}},\ }\bibfield  {title} {\bibinfo
  {title} {Multinucleon transfer mechanism in
  $^{160}\mathrm{Gd}+^{186}\mathrm{W}$ collisions in stochastic mean-field
  theory},\ }\href {https://doi.org/10.1103/PhysRevC.108.054605} {\bibfield
  {journal} {\bibinfo  {journal} {Phys. Rev. C}\ }\textbf {\bibinfo {volume}
  {108}},\ \bibinfo {pages} {054605} (\bibinfo {year}
  {2023}{\natexlab{b}})}\BibitemShut {NoStop}%
\bibitem [{\citenamefont {{Kayaalp, A.}}\ \emph {et~al.}(2024)\citenamefont
  {{Kayaalp, A.}}, \citenamefont {{Ocal, S. E.}}, \citenamefont {{Yaprakli,
  B.}}, \citenamefont {{Arik, M.}}, \citenamefont {{Ayik, S.}}, \citenamefont
  {{Yilmaz, O.}},\ and\ \citenamefont {{Umar, A. S.}}}]{kayaalp2024}%
  \BibitemOpen
  \bibfield  {author} {\bibinfo {author} {\bibnamefont {{Kayaalp, A.}}},
  \bibinfo {author} {\bibnamefont {{Ocal, S. E.}}}, \bibinfo {author}
  {\bibnamefont {{Yaprakli, B.}}}, \bibinfo {author} {\bibnamefont {{Arik,
  M.}}}, \bibinfo {author} {\bibnamefont {{Ayik, S.}}}, \bibinfo {author}
  {\bibnamefont {{Yilmaz, O.}}},\ and\ \bibinfo {author} {\bibnamefont {{Umar,
  A. S.}}},\ }\bibfield  {title} {\bibinfo {title} {A theoretical study on
  quasifission and fusion--fission processes in heavy-ion collisions},\ }\href
  {https://doi.org/10.1140/epja/s10050-024-01303-9} {\bibfield  {journal}
  {\bibinfo  {journal} {Eur. Phys. J. A}\ }\textbf {\bibinfo {volume} {60}},\
  \bibinfo {pages} {79} (\bibinfo {year} {2024})}\BibitemShut {NoStop}%
\bibitem [{\citenamefont {Kratz}\ \emph {et~al.}(2013)\citenamefont {Kratz},
  \citenamefont {Sch\"adel},\ and\ \citenamefont {G\"aggeler}}]{kratz2013}%
  \BibitemOpen
  \bibfield  {author} {\bibinfo {author} {\bibfnamefont {J.~V.}\ \bibnamefont
  {Kratz}}, \bibinfo {author} {\bibfnamefont {M.}~\bibnamefont {Sch\"adel}},\
  and\ \bibinfo {author} {\bibfnamefont {H.~W.}\ \bibnamefont {G\"aggeler}},\
  }\bibfield  {title} {\bibinfo {title} {Reexamining the heavy-ion reactions
  $^{238}\mathrm{U}+{}^{238}\mathrm{U}$ and
  $^{238}\mathrm{U}+{}^{248}\mathrm{Cm}$ and actinide production close to the
  barrier},\ }\href {https://doi.org/10.1103/PhysRevC.88.054615} {\bibfield
  {journal} {\bibinfo  {journal} {Phys. Rev. C}\ }\textbf {\bibinfo {volume}
  {88}},\ \bibinfo {pages} {054615} (\bibinfo {year} {2013})}\BibitemShut
  {NoStop}%
\bibitem [{\citenamefont {Sch\"adel}\ \emph {et~al.}(1982)\citenamefont
  {Sch\"adel}, \citenamefont {Br\"uchle}, \citenamefont {G\"aggeler},
  \citenamefont {Kratz}, \citenamefont {S\"ummerer}, \citenamefont {Wirth},
  \citenamefont {Herrmann}, \citenamefont {Stakemann}, \citenamefont {Tittel},
  \citenamefont {Trautmann}, \citenamefont {Nitschke}, \citenamefont {Hulet},
  \citenamefont {Lougheed}, \citenamefont {Hahn},\ and\ \citenamefont
  {Ferguson}}]{schaedel1982}%
  \BibitemOpen
  \bibfield  {author} {\bibinfo {author} {\bibfnamefont {M.}~\bibnamefont
  {Sch\"adel}}, \bibinfo {author} {\bibfnamefont {W.}~\bibnamefont
  {Br\"uchle}}, \bibinfo {author} {\bibfnamefont {H.}~\bibnamefont
  {G\"aggeler}}, \bibinfo {author} {\bibfnamefont {J.~V.}\ \bibnamefont
  {Kratz}}, \bibinfo {author} {\bibfnamefont {K.}~\bibnamefont {S\"ummerer}},
  \bibinfo {author} {\bibfnamefont {G.}~\bibnamefont {Wirth}}, \bibinfo
  {author} {\bibfnamefont {G.}~\bibnamefont {Herrmann}}, \bibinfo {author}
  {\bibfnamefont {R.}~\bibnamefont {Stakemann}}, \bibinfo {author}
  {\bibfnamefont {G.}~\bibnamefont {Tittel}}, \bibinfo {author} {\bibfnamefont
  {N.}~\bibnamefont {Trautmann}}, \bibinfo {author} {\bibfnamefont {J.~M.}\
  \bibnamefont {Nitschke}}, \bibinfo {author} {\bibfnamefont {E.~K.}\
  \bibnamefont {Hulet}}, \bibinfo {author} {\bibfnamefont {R.~W.}\ \bibnamefont
  {Lougheed}}, \bibinfo {author} {\bibfnamefont {R.~L.}\ \bibnamefont {Hahn}},\
  and\ \bibinfo {author} {\bibfnamefont {R.~L.}\ \bibnamefont {Ferguson}},\
  }\bibfield  {title} {\bibinfo {title} {Actinide {P}roduction in {C}ollisions
  of $^{238}\mathrm{U}$ with $^{248}\mathrm{Cm}$},\ }\href
  {https://doi.org/10.1103/PhysRevLett.48.852} {\bibfield  {journal} {\bibinfo
  {journal} {Phys. Rev. Lett.}\ }\textbf {\bibinfo {volume} {48}},\ \bibinfo
  {pages} {852} (\bibinfo {year} {1982})}\BibitemShut {NoStop}%
\bibitem [{\citenamefont {Kratz}\ \emph {et~al.}(2015)\citenamefont {Kratz},
  \citenamefont {Loveland},\ and\ \citenamefont {Moody}}]{kratz2015}%
  \BibitemOpen
  \bibfield  {author} {\bibinfo {author} {\bibfnamefont {J.~V.}\ \bibnamefont
  {Kratz}}, \bibinfo {author} {\bibfnamefont {W.}~\bibnamefont {Loveland}},\
  and\ \bibinfo {author} {\bibfnamefont {K.~J.}\ \bibnamefont {Moody}},\
  }\bibfield  {title} {\bibinfo {title} {Syntheses of transuranium isotopes
  with atomic numbers ${Z}\le 103$ in multi-nucleon transfer reactions},\
  }\href {https://doi.org/10.1016/j.nuclphysa.2015.06.004} {\bibfield
  {journal} {\bibinfo  {journal} {Nucl. Phys. A}\ }\textbf {\bibinfo {volume}
  {944}},\ \bibinfo {pages} {117} (\bibinfo {year} {2015})}\BibitemShut
  {NoStop}%
\bibitem [{\citenamefont {Peng}\ and\ \citenamefont {Feng}(2022)}]{peng2022}%
  \BibitemOpen
  \bibfield  {author} {\bibinfo {author} {\bibfnamefont {C.}~\bibnamefont
  {Peng}}\ and\ \bibinfo {author} {\bibfnamefont {Z.-Q.}\ \bibnamefont
  {Feng}},\ }\bibfield  {title} {\bibinfo {title} {Production of neutron-rich
  heavy nuclei around ${N}=162$ in multinucleon transfer reactions},\ }\href
  {https://doi.org/10.1140/epja/s10050-022-00819-2} {\bibfield  {journal}
  {\bibinfo  {journal} {Eur. Phys. J. A}\ }\textbf {\bibinfo {volume} {58}},\
  \bibinfo {pages} {162} (\bibinfo {year} {2022})}\BibitemShut {NoStop}%
\bibitem [{\citenamefont {Zhao}\ \emph {et~al.}(2023)\citenamefont {Zhao},
  \citenamefont {Bao},\ and\ \citenamefont {Zhang}}]{zhao2023}%
  \BibitemOpen
  \bibfield  {author} {\bibinfo {author} {\bibfnamefont {T.~L.}\ \bibnamefont
  {Zhao}}, \bibinfo {author} {\bibfnamefont {X.~J.}\ \bibnamefont {Bao}},\ and\
  \bibinfo {author} {\bibfnamefont {H.~F.}\ \bibnamefont {Zhang}},\ }\bibfield
  {title} {\bibinfo {title} {Exploring the optimal way to produce ${Z}$=100-106
  neutron-rich nuclei},\ }\href {https://doi.org/10.1103/PhysRevC.108.024602}
  {\bibfield  {journal} {\bibinfo  {journal} {Phys. Rev. C}\ }\textbf {\bibinfo
  {volume} {108}},\ \bibinfo {pages} {024602} (\bibinfo {year}
  {2023})}\BibitemShut {NoStop}%
\bibitem [{\citenamefont {{Hannes Risken}}\ and\ \citenamefont {{Till
  Frank}}(1996)}]{risken1996}%
  \BibitemOpen
  \bibfield  {author} {\bibinfo {author} {\bibnamefont {{Hannes Risken}}}\ and\
  \bibinfo {author} {\bibnamefont {{Till Frank}}},\ }\href
  {https://doi.org/10.1007/978-3-642-61544-3} {\emph {\bibinfo {title} {The
  {F}okker--{P}lanck {E}quation}}}\ (\bibinfo  {publisher} {Springer--Verlag},\
  \bibinfo {address} {Berlin},\ \bibinfo {year} {1996})\BibitemShut {NoStop}%
\bibitem [{\citenamefont {Schr\"oder}\ \emph {et~al.}(1981)\citenamefont
  {Schr\"oder}, \citenamefont {Huizenga},\ and\ \citenamefont
  {Randrup}}]{schroder1981}%
  \BibitemOpen
  \bibfield  {author} {\bibinfo {author} {\bibfnamefont {W.~U.}\ \bibnamefont
  {Schr\"oder}}, \bibinfo {author} {\bibfnamefont {J.~R.}\ \bibnamefont
  {Huizenga}},\ and\ \bibinfo {author} {\bibfnamefont {J.}~\bibnamefont
  {Randrup}},\ }\bibfield  {title} {\bibinfo {title} {Correlated mass and
  charge transport induced by statistical nucleon exchange in damped nuclear
  reactions},\ }\href {https://doi.org/10.1016/0370-2693(81)90924-2} {\bibfield
   {journal} {\bibinfo  {journal} {Phys. Lett. B}\ }\textbf {\bibinfo {volume}
  {98}},\ \bibinfo {pages} {355} (\bibinfo {year} {1981})}\BibitemShut
  {NoStop}%
\bibitem [{\citenamefont {Merchant}\ and\ \citenamefont
  {N\"orenberg}(1981)}]{merchant1981}%
  \BibitemOpen
  \bibfield  {author} {\bibinfo {author} {\bibfnamefont {A.~C.}\ \bibnamefont
  {Merchant}}\ and\ \bibinfo {author} {\bibfnamefont {W.}~\bibnamefont
  {N\"orenberg}},\ }\bibfield  {title} {\bibinfo {title} {Neutron and proton
  diffusion in heavy--ion collisions},\ }\href
  {https://doi.org/10.1016/0370-2693(81)90844-3} {\bibfield  {journal}
  {\bibinfo  {journal} {Phys. Lett. B}\ }\textbf {\bibinfo {volume} {104}},\
  \bibinfo {pages} {15} (\bibinfo {year} {1981})}\BibitemShut {NoStop}%
\bibitem [{\citenamefont {Gardiner}(1991)}]{gardiner1991}%
  \BibitemOpen
  \bibfield  {author} {\bibinfo {author} {\bibfnamefont {C.~W.}\ \bibnamefont
  {Gardiner}},\ }\href@noop {} {\emph {\bibinfo {title} {Quantum {N}oise}}}\
  (\bibinfo  {publisher} {Springer--{V}erlag},\ \bibinfo {address} {Berlin},\
  \bibinfo {year} {1991})\BibitemShut {NoStop}%
\bibitem [{\citenamefont {Weiss}(1999)}]{weiss1999}%
  \BibitemOpen
  \bibfield  {author} {\bibinfo {author} {\bibfnamefont {U.}~\bibnamefont
  {Weiss}},\ }\href@noop {} {\emph {\bibinfo {title} {Quantum {D}issipative
  {S}ystems}}},\ \bibinfo {edition} {2nd}\ ed.\ (\bibinfo  {publisher} {World
  {S}cientific},\ \bibinfo {address} {Singapore},\ \bibinfo {year}
  {1999})\BibitemShut {NoStop}%
\bibitem [{\citenamefont {N\"orenberg}(1981)}]{norenberg1981}%
  \BibitemOpen
  \bibfield  {author} {\bibinfo {author} {\bibfnamefont {W.}~\bibnamefont
  {N\"orenberg}},\ }\bibfield  {title} {\bibinfo {title} {Memory effects in the
  energy dissipation for slow collective nuclear motion},\ }\href
  {https://doi.org/10.1016/0370-2693(81)90570-0} {\bibfield  {journal}
  {\bibinfo  {journal} {Phys. Lett. B}\ }\textbf {\bibinfo {volume} {104}},\
  \bibinfo {pages} {107} (\bibinfo {year} {1981})}\BibitemShut {NoStop}%
\bibitem [{\citenamefont {Randrup}(1979)}]{randrup1979}%
  \BibitemOpen
  \bibfield  {author} {\bibinfo {author} {\bibfnamefont {J.}~\bibnamefont
  {Randrup}},\ }\bibfield  {title} {\bibinfo {title} {Theory of
  transfer-induced transport in nuclear collisions},\ }\href
  {https://doi.org/10.1016/0375-9474(79)90271-9} {\bibfield  {journal}
  {\bibinfo  {journal} {Nucl. Phys. A}\ }\textbf {\bibinfo {volume} {327}},\
  \bibinfo {pages} {490} (\bibinfo {year} {1979})}\BibitemShut {NoStop}%
\bibitem [{\citenamefont {Merchant}\ and\ \citenamefont
  {N\"orenberg}(1982)}]{merchant1982}%
  \BibitemOpen
  \bibfield  {author} {\bibinfo {author} {\bibfnamefont {A.~C.}\ \bibnamefont
  {Merchant}}\ and\ \bibinfo {author} {\bibfnamefont {W.}~\bibnamefont
  {N\"orenberg}},\ }\bibfield  {title} {\bibinfo {title} {Microscopic transport
  theory of heavy-ion collisions},\ }\href {https://doi.org/10.1007/bf01415880}
  {\bibfield  {journal} {\bibinfo  {journal} {Z. Phys. A}\ }\textbf {\bibinfo
  {volume} {308}},\ \bibinfo {pages} {315} (\bibinfo {year}
  {1982})}\BibitemShut {NoStop}%
\bibitem [{\citenamefont {Charity}(2008)}]{charity2008}%
  \BibitemOpen
  \bibfield  {author} {\bibinfo {author} {\bibfnamefont {R.}~\bibnamefont
  {Charity}},\ }\href
  {https://inis.iaea.org/search/search.aspx?orig_q=RN:40048000} {\emph
  {\bibinfo {title} {{G}{EMINI: A} code to simulate the decay of a compound
  nucleus by a series of binary decays}}},\ \bibinfo {type} {Tech. Rep.}\
  (\bibinfo  {institution} {International {A}tomic {E}nergy {A}gency
  ({IAEA})},\ \bibinfo {year} {2008})\ \bibinfo {note}
  {{INDC(NDS)}--0530}\BibitemShut {NoStop}%
\bibitem [{\citenamefont {Hauser}\ and\ \citenamefont
  {Feshbach}(1952)}]{hauser1952}%
  \BibitemOpen
  \bibfield  {author} {\bibinfo {author} {\bibfnamefont {W.}~\bibnamefont
  {Hauser}}\ and\ \bibinfo {author} {\bibfnamefont {H.}~\bibnamefont
  {Feshbach}},\ }\bibfield  {title} {\bibinfo {title} {The {I}nelastic
  {S}cattering of {N}eutrons},\ }\href {https://doi.org/10.1103/PhysRev.87.366}
  {\bibfield  {journal} {\bibinfo  {journal} {Phys. Rev.}\ }\textbf {\bibinfo
  {volume} {87}},\ \bibinfo {pages} {366} (\bibinfo {year} {1952})}\BibitemShut
  {NoStop}%
\bibitem [{\citenamefont {Moretto}(1975)}]{moretto1975}%
  \BibitemOpen
  \bibfield  {author} {\bibinfo {author} {\bibfnamefont {L.~G.}\ \bibnamefont
  {Moretto}},\ }\bibfield  {title} {\bibinfo {title} {Statistical emission of
  large fragments: {A} general theoretical approach},\ }\href
  {https://doi.org/10.1016/0375-9474(75)90632-6} {\bibfield  {journal}
  {\bibinfo  {journal} {Nucl. Phys. A}\ }\textbf {\bibinfo {volume} {247}},\
  \bibinfo {pages} {211} (\bibinfo {year} {1975})}\BibitemShut {NoStop}%
\bibitem [{\citenamefont {Moretto}\ and\ \citenamefont
  {Wozniak}(1988)}]{moretto1988}%
  \BibitemOpen
  \bibfield  {author} {\bibinfo {author} {\bibfnamefont {L.~G.}\ \bibnamefont
  {Moretto}}\ and\ \bibinfo {author} {\bibfnamefont {G.~J.}\ \bibnamefont
  {Wozniak}},\ }\bibfield  {title} {\bibinfo {title} {The role of the compound
  nucleus in complex fragment emission at low and intermediate energies},\
  }\href {https://doi.org/10.1016/0146-6410(88)90036-1} {\bibfield  {journal}
  {\bibinfo  {journal} {Prog. Part. Nucl. Phys.}\ }\textbf {\bibinfo {volume}
  {21}},\ \bibinfo {pages} {401} (\bibinfo {year} {1988})}\BibitemShut
  {NoStop}%
\bibitem [{\citenamefont {Blatt}\ and\ \citenamefont
  {Weisskopf}(1979)}]{blatt1979}%
  \BibitemOpen
  \bibfield  {author} {\bibinfo {author} {\bibfnamefont {J.~M.}\ \bibnamefont
  {Blatt}}\ and\ \bibinfo {author} {\bibfnamefont {V.~F.}\ \bibnamefont
  {Weisskopf}},\ }\href {https://doi.org/10.1007/978-1-4612-9959-2} {\emph
  {\bibinfo {title} {Theoretical {N}uclear {P}hysics}}}\ (\bibinfo  {publisher}
  {Springer New York, NY},\ \bibinfo {year} {1979})\BibitemShut {NoStop}%
\bibitem [{\citenamefont {{David J. Kedziora}}\ and\ \citenamefont {{C\'edric
  Simenel}}(2010)}]{kedziora2010}%
  \BibitemOpen
  \bibfield  {author} {\bibinfo {author} {\bibnamefont {{David J. Kedziora}}}\
  and\ \bibinfo {author} {\bibnamefont {{C\'edric Simenel}}},\ }\bibfield
  {title} {\bibinfo {title} {{N}ew inverse quasifission mechanism to produce
  neutron-rich transfermium nuclei},\ }\href
  {https://doi.org/10.1103/PhysRevC.81.044613} {\bibfield  {journal} {\bibinfo
  {journal} {Phys. Rev. C}\ }\textbf {\bibinfo {volume} {81}},\ \bibinfo
  {pages} {044613} (\bibinfo {year} {2010})}\BibitemShut {NoStop}%
\bibitem [{\citenamefont {Umar}\ \emph {et~al.}(1991)\citenamefont {Umar},
  \citenamefont {Strayer}, \citenamefont {Wu}, \citenamefont {Dean},\ and\
  \citenamefont {G\"u\c{c}l\"u}}]{umar1991a}%
  \BibitemOpen
  \bibfield  {author} {\bibinfo {author} {\bibfnamefont {A.~S.}\ \bibnamefont
  {Umar}}, \bibinfo {author} {\bibfnamefont {M.~R.}\ \bibnamefont {Strayer}},
  \bibinfo {author} {\bibfnamefont {J.~S.}\ \bibnamefont {Wu}}, \bibinfo
  {author} {\bibfnamefont {D.~J.}\ \bibnamefont {Dean}},\ and\ \bibinfo
  {author} {\bibfnamefont {M.~C.}\ \bibnamefont {G\"u\c{c}l\"u}},\ }\bibfield
  {title} {\bibinfo {title} {{N}uclear {H}artree-{F}ock calculations with
  splines},\ }\href {https://doi.org/10.1103/PhysRevC.44.2512} {\bibfield
  {journal} {\bibinfo  {journal} {Phys. Rev. C}\ }\textbf {\bibinfo {volume}
  {44}},\ \bibinfo {pages} {2512} (\bibinfo {year} {1991})}\BibitemShut
  {NoStop}%
\bibitem [{\citenamefont {Umar}\ and\ \citenamefont
  {Oberacker}(2006)}]{umar2006c}%
  \BibitemOpen
  \bibfield  {author} {\bibinfo {author} {\bibfnamefont {A.~S.}\ \bibnamefont
  {Umar}}\ and\ \bibinfo {author} {\bibfnamefont {V.~E.}\ \bibnamefont
  {Oberacker}},\ }\bibfield  {title} {\bibinfo {title} {{T}hree-dimensional
  unrestricted time-dependent {H}artree-{F}ock fusion calculations using the
  full {S}kyrme interaction},\ }\href
  {https://doi.org/10.1103/PhysRevC.73.054607} {\bibfield  {journal} {\bibinfo
  {journal} {Phys. Rev. C}\ }\textbf {\bibinfo {volume} {73}},\ \bibinfo
  {pages} {054607} (\bibinfo {year} {2006})}\BibitemShut {NoStop}%
\bibitem [{\citenamefont {{Ka--Hae Kim}}\ \emph {et~al.}(1997)\citenamefont
  {{Ka--Hae Kim}}, \citenamefont {{Takaharu Otsuka}},\ and\ \citenamefont
  {{Paul Bonche}}}]{kim1997}%
  \BibitemOpen
  \bibfield  {author} {\bibinfo {author} {\bibnamefont {{Ka--Hae Kim}}},
  \bibinfo {author} {\bibnamefont {{Takaharu Otsuka}}},\ and\ \bibinfo {author}
  {\bibnamefont {{Paul Bonche}}},\ }\bibfield  {title} {\bibinfo {title}
  {{T}hree-dimensional {TDHF} calculations for reactions of unstable nuclei},\
  }\href {https://doi.org/10.1088/0954-3899/23/10/014} {\bibfield  {journal}
  {\bibinfo  {journal} {J. Phys. G: Nucl. Part. Phys.}\ }\textbf {\bibinfo
  {volume} {23}},\ \bibinfo {pages} {1267} (\bibinfo {year}
  {1997})}\BibitemShut {NoStop}%
\end{thebibliography}%

\end{document}